\pdfoutput=1
\documentclass{emulateapj}
\usepackage{xspace}
\usepackage{amsmath}
\usepackage{hyperref}
\usepackage{framed} 
\usepackage{txfonts}
\usepackage{epstopdf} 
\def\Ht{{\rm H}}
\def\H2{{{\rm H}_2}}
\def\HI{{\rm H\,I}}
\def\HII{{\rm H\,II}}
\def\GI{{\rm He\,I}}
\def\GII{{\rm He\,II}}
\def\GIII{{\rm He\,III}}
\def\Hp{{{\rm H}_2^+}}
\def\Hm{{{\rm H}^-}}
\def\MH{\H2}

\def\Msun{\, M_{\odot}}

\def\Sgas{\Sigma_{\rm H}}
\def\Sntr{\Sigma_{\HI+\H2}}
\def\Stot{\Sigma_{\HI+\HII+\H2}}
\def\Smol{\Sigma_\H2}
\def\Shi{\Sigma_\HI}
\def\Shii{\Sigma_\HII}
\def\Ssfr{\Sigma_{\rm SFR}}

\def\D{D_{\rm MW}}
\def\U{U_{\rm MW}}
\def\AM{{\HI\rightarrow\H2}}

\def\SD{S_{\rm D}}
\def\GA{\Gamma_{\rm A}}
\def\GB{\Gamma_{\rm B}}
\def\GC{\Gamma_{\rm C}}
\def\GE{\Gamma_{\rm E}}
\def\GD{\Gamma_{\rm D}}
\def\GLW{\Gamma_{\rm LW}}

\def\dim#1{\mbox{\,#1}}

\def\hide#1{}

\begin{document}

%=================================================
\title{Environmental Dependence of the Kennicutt-Schmidt Relation in Galaxies}
%=================================================

\author{Nickolay Y.\
  Gnedin\altaffilmark{1,2,3} and Andrey
  V. Kravtsov\altaffilmark{2,3,4}}  
\altaffiltext{1}{Particle Astrophysics Center, 
Fermi National Accelerator Laboratory, Batavia, IL 60510, USA; gnedin@fnal.gov}
\altaffiltext{2}{Kavli Institute for Cosmological Physics and Enrico
  Fermi Institute, The University of Chicago, Chicago, IL 60637 USA;
  andrey@oddjob.uchicago.edu} 
\altaffiltext{3}{Department of Astronomy \& Astrophysics, The
  University of Chicago, Chicago, IL 60637 USA} 
\altaffiltext{4}{Enrico Fermi Institute, The University of Chicago,
Chicago, IL 60637}

\begin{abstract}
We present a detailed description of a phenomenological $\H2$
formation model and local star formation prescription based on the
density of molecular (rather than total) gas. Such approach allows us
to avoid the arbitrary density and temperature thresholds typically
used in star formation recipes.  We present results of the model based
on realistic cosmological simulations of high-$z$ galaxy formation for
a grid of numerical models with varied dust-to-gas ratios and
interstellar far UV (FUV) fluxes. Our results show that both the
atomic-to-molecular transition on small, $\sim 10\dim{pc}$ scales and
the Kennicutt-Schmidt (KS) relation on $\sim\dim{kpc}$ scales are
sensititive to the dust-to-gas ratio and the FUV flux. The
atomic-to-molecular transition as a function of gas density or column
density has a large scatter but is rather sharp and shifts to higher
densities with decreasing dust-to-gas ratio and/or increasing FUV flux.
Consequently, star formation is concentrated to higher gas surface
density regions, resulting in steeper slope and lower amplitude of the
KS relation at a given $\Sgas$, in less dusty and/or higher FUV flux
environments. These trends should have a particularly strong effect on
the evolution of low-mass, low surface brightness galaxies which
typically have low dust content and anemic star formation, but are also likely to be important for evolution of the Milky Way-sized systems.  We
parameterize the dependencies observed in our simulations in
convenient fitting formulae, which can be used to model the dependence
of the KS relation on the dust-to-gas ratio and FUV flux in
semi-analytic models and in cosmological simulations that do not
include radiative transfer and $\H2$ formation.
\end{abstract}


%----------------------
\section{Introduction}
\label{sec:intro}
%----------------------

Conversion of gas into stars is one of the major sources of
uncertainty in modeling formation of galaxies.  This uncertainty
reflects our incomplete understanding of the process of star
formation both locally and on global scales.  Traditionally, star
formation is included in cosmological simulations and simulations of
isolated galaxies by using simple phenomenological prescriptions that
relate local rate of star formation to the local density of gas, with
some additional criteria such as temperature and density thresholds
for the gas to be eligible for star formation.  The parameters of
these prescriptions are chosen so that the empirical power law
relation between the {\it surface density} of star formation, $\Ssfr$,
and surface density of (hydrogen) gas averaged on kpc scales, $\Sgas$,
$\Ssfr\propto \Sgas^n$ with $n\approx 1 - 1.4$,
\citep{schmidt59,sfr:k98a,sfr:blwb08} observed in $z\approx 0$
galaxies is reproduced \citep[see, e.g.,][for a recent overview]{schaye_dallavecchia08}.

However, both theoretical considerations and observational evidence
indicate that such approach may miss some important environmental
trends. For example, relation between the local star formation recipe
and the large-scale Kennicutt-Schmidt (KS) relation is not trivial and
depends on the density and thermal structure of the interstellar
medium
\citep[ISM,][]{sims:k03,tassis07,wada_norman07,robertson_kravtsov08,schaye_dallavecchia08,saitoh_etal08}.
This is because for a given large-scale gas surface density the
fraction of dense, star forming gas is determined by the gas density
distribution function, which, in turn, depends on the thermal state of
the ISM \citep{wada_norman01,robertson_kravtsov08}. For the same
reason, the global rate of star formation may be controlled by the
rate with which dense gas is formed by the ISM, rather then by the
assumed local efficiency of the gas \citep{saitoh_etal08}. This
implies that star formation parameters tuned to reproduce the
empirical KS relation in one situation \citep[e.g., in controlled
simulations of isolated
disks][]{springel_hernquist03,schaye_dallavecchia08} may not reproduce
this relation in galaxies with significantly different ISM density
distributions.

In addition, there is a growing observational evidence that the
KS relation is more complex than previously thought
\citep{heyer_etal04,boissier_etal03,sfr:blwb08}. For example, instead
of a well-defined surface density threshold at low $\Sgas$ below which
$\Ssfr$ drops to zero \citep{martin_kennicutt01}, observations
indicate continuous relation between star formation rate and gas
surface densities \citep{boissier_etal07} down to small $\Sgas$,
albeit with a steeper slope \citep[e.g.,][]{sfr:blwb08}. Likewise,
studies of individual dwarf galaxies, which typically have low gas
surface densities ($\Sgas\lesssim 10-20\Msun\dim{pc}^{-2}$)
throughout their disks, show that the KS relation in such
galaxies is generally characterized by a considerably steeper slope,
$n\approx 2-4$, than the canonical value of 1.4
\citep{heyer_etal04,sfr:blwb08,verley_etal10}. Moreover,
recent detailed study of the global star formation relation by
\citet{sfr:blwb08} shows that a single power law is in general a poor
description of the KS relation over the entire range of surface
densities. Instead, the slope of the $\Ssfr-\Sgas$ relation may vary
from the steep values of $n\approx 2-4$ at $\Sgas\la 10
\Msun\dim{pc}^{-2}$ to linear $n\approx 1$ at $\Sgas\sim 10-100
\Msun\dim{pc}^{-2}$ and then possibly steepening again to $n\approx
1.5-2$ at $\Sgas\gtrsim 100\Msun\dim{pc}^{-2}$.

Finally, the growing evidence indicates that in high-redshift galaxies
($z\gtrsim 3$) the KS relation is significantly steeper and has an
order of magnitude lower amplitude at $\Ssfr\lesssim
100\Msun\dim{pc}^{-2}$ \citep[][see also Fig. 3 in
\citeauthor{ng:gk10a}
\citeyear{ng:gk10a}]{sfr:wc06,rafelski_etal10}.

This complex behavior of the star formation rate density with the
density of the neutral gas ($\HI$+$\H2$) can be understood if star
formation occurs only in the molecular gas
\citep{robertson_kravtsov08,ng:gtk09,krumholz_etal09,pelupessy_popadopoulos09,ng:gk10a}. Indeed,
detailed observations of nearby galaxies show that star formation
correlates most strongly with the molecular gas
\citep[e.g.,][]{wong_blitz02,sfr:blwb08}, especially with the densest
gas traced by HCN emission \citep{gao_solomon04,wu_etal05}, while it
only correlates weakly, if at all, with the density of atomic gas
\citep{wong_blitz02,kennicutt_etal07,sfr:blwb08}. We can thus expect
that the relationship between the star formation rate density and gas
density $\Sgas=\Smol+\Shi$ (the KS relation) varies depending on the
molecular fraction of the gas $f_\H2=\Smol/\Sgas$.

Several factors may control the molecular fraction in the gas on
different spatial scales. On small scales of individual molecular
complexes it is primarily the cosmic dust abundance and the
interstellar FUV radiation that control the atomic-to-molecular
transition \citep[e.g.,][see \citeauthor{stahler_palla05}
\citeyear{stahler_palla05} for pedagogical
review]{elmegreen93,sfr:kmt08}. On larger ($\sim\dim{kpc}$) scales the
fraction of dense, molecular gas in a patch of gas of a given $\Sgas$
is expected to depend on the density distribution of gas in that patch
\citep[e.g.,][]{elmegreen02}. The density distribution itself depends
on thermodynamics of gas \citep[see, e.g.,][]{robertson_kravtsov08}
and metallicity, as more metal rich gas may be more efficient in
building regions of higher densities via radiative shocks arising in
the highly turbulent medium of gaseous disks. The density PDF should
also reflect the global dynamics of gas in galactic disks in
general. For example, spiral density wave will compress the gas
facilitating its cooling and conversion of atomic gas into molecular
form. Likewise, large-scale instabilities seed the turbulence in the
disk that can shape the global density PDF
\citep{wada_norman01,elmegreen02,sims:k03,sfr:km05}.

Although observational studies of environmental dependence of the KS
relation on gas metallicity, interstellar FUV radiation, and other
properties of galaxies are in their early stages
\citep[e.g.,][]{sfr:blwb08,sfr:kept09,rafelski_etal10}, it is clear
that such strong dependences can have important implications for our
understanding of galaxy evolution \cite[see discussion
in][]{ng:gk10a}. For example, given that observations
indicate that star formation in low-metallicity, high-UV flux
environments of high-redshift galaxies is concentrated to
significantly higher gas surface densities
\citep{sfr:wc06,rafelski_etal10}, stars in these galaxies should be
confined to the high surface density regions and should therefore be
more resistant against dynamical heating in mergers. At the same time,
the longer gas consumption time scales in lower density regions of
high-$z$ gaseous disks along with high accretion rate would keep them
gas rich and more resilient to mergers as well
\citep[e.g..][]{robertson_etal04,robertson_etal06,springel_hernquist05}. This
can help to resolve one of the major puzzles of hierarchical galaxy
formation: prevalence of thin disks at low redshifts in the face of
high merger rates at high redshifts. 

It is thus important to explore potential effects and implications of
the enviromental dependence of the KS relation for the evolution of
galaxies. However, to capture the key physics responsible for this
dependence in cosmological simulations of galaxy formation is
challenging, because this requires high spatial resolution to model
dynamics of interstellar medium in the hierarchically forming
galaxies, 3D radiative transfer to model local UV radiation flux, and
formation of molecular hydrogen. The latter is mediated by dust grains
which catalyze H$_2$ formation and provide the initial key shielding
from interstellar FUV radiation. This shielding allows build-up of
molecular fraction sufficient for H$_2$ self-shielding, which in turn
shapes the sharp transition of atomic to molecular gas. 

Although fully self-consistent modeling of dust chemistry and H$_2$
formation is still far beyond reach, phenomenological model capturing
the essential metallicity and UV flux dependence of molecular fraction
can be used to model H$_2$ in self-consistent, high-resolution
cosmological simulations \citep[][]{ng:gtk09,ng:gk10a}. In
this study we present a detailed description of such H$_2$ formation
model and local star formation prescription based on the density of
molecular (rather than total) gas. We present results for a grid of
numerical models with varied dust-to-gas ratios and interstellar FUV
radiation fluxes and explore the dependence of atomic-to-molecular
transition on small, molecular cloud scales, on these variables and
the effect this dependence has on the Kennicutt-Schmidt relation on
large $\sim\dim{kpc}$ scales. We parameterize the dependencies
observed in our simulations in convenient fitting formulae, which can
be used to model the metallicity and UV flux dependence of the KS
relation in semi-analytic models and in cosmological simulations that
do not include radiative transfer and $\H2$ formation.

%---------------------
\section{Simulations}
\label{sec:sims}
%---------------------

For our tests we use the simulation of galaxy formation described in
\citet{ng:gtk09}.  The simulation was run with Adaptive Refinement
Tree (ART) code \citep{kravtsov99,kravtsov_etal02,rudd_etal08} and
follows a Lagrangian region corresponding to five virial radii of a
system, which evolves into a typical halo of an $L_{\ast}$ galaxy
($M\approx 10^{12}\Msun$) at $z=0$. The mass resolution in the
high-resolution Lagrangian region is $1.3\times10^6\Msun$ in dark
matter and mass resolution in baryon that varies from $\sim 10^3\Msun$
to $\sim 10^6\Msun$ depending on the cell size and density. The
simulation reaches peak spatial resolution of $260$ comoving pc
($65\dim{pc}$ in physical units at $z=3$). The Lagrangian region is
embedded into a cubic volume of $6h^{-1}$ comoving Mpc on a side to
model the tidal forces from the surrounding structures properly, but
this outer region is resolved only coarsely with a uniform $64^3$
grid.

The cosmological simulation follows collapse of dark matter and gas
self-consistently. The heating and cooling of gas is followed as well,
so that gas can dissipate the energy it gains during collapse and sink
to the center of its parent halo. Our simulations include 3D radiative
transfer (RT) of UV radiation from individual stellar particles formed
during the course of the simulation using the OTVET approximation
\citep{ng:ga01}. Inclusion of the RT is important because the local UV
flux can set ionization and heating balance of gas and influence the
abundance of molecular hydrogen, as we descibe below and in the
Appendix. Unlike the IGM after reionization, which can be assumed
optically thin to ionizing radiation, the dense ISM gas of simulated
galaxies may well be opaque to ionizing photons of all but the nearest
stars.

The simulations incorporate non-equilibrium chemical network of
hydrogen and helium and non-equilibrium cooling and heating rates,
which make use of the local abundance of atomic, molecular, and ionic
species and UV intensity.  This network includes formation of
molecular hydrogen both in the primordial phase and on dust
grains. The abundances of the relevant atomic and molecular species
are therefore followed self-consistently during the course of the
simulation. The heating and cooling terms in the equation for the
internal energy include all of the terms normally included in the
simulations of first stars and in the ISM models, including cooling on
metals. We describe all included reactions and heating/cooling
processes in Appendix.

The model also accounts both for self-shielding of $\H2$ from the
dissociating FUV radiation and the shielding provided by the
interstellar dust using phenomenological prescriptions for shielding
factors. The details of the model are presented in the Appendix. Our
model is calibrated against the observed column density dependence of
atomic and molecular gas fractions in the Milky Way, LMC, and SMC
(see Appendix). In particular, the model reproduces the metallicity
dependence of the column density of the sharp transition from the
atomic to fully molecular gas observed in the MW, LMC, and SMC.

In order to investigate the environmental dependence of the star
formation rate in the simulations, we perform a series of controlled
test simulations. For each of these tests, we fix the dust-to-gas
ratio in the $\H2$ model and normalization of the emissivity of
stellar particles at $1000\ \AA$ to constant values and run the
simulations for a significant period of time. 

We explore a grid of values of dust-to-gas ratio $\D$ from $10^{-3}$
to $1.0$ relative to the Milky Way value. The variable $\D$ scales the
H$_2$ on dust formation rate coefficient $R_D$ and the absorption
cross-section of dust in the Lyman-Werner band $\sigma_{\rm LW}$ 
to the values characteristic for the Milky Way:
\begin{equation}
  R_D \equiv \D R_0;\ \ \ \sigma_{\rm LW} \equiv \D\sigma_0,
  \label{eq:pardefs}
\end{equation}
where
$R_0=3.5\times 10^{-17}\rm\ cm^3\, s^{-1}$ \citep{ism:wthk08} and
$\sigma_0=2\times 10^{-21}\rm\ cm^2$ \citep{ism:db96,ism:gm07a},
respectively. 

The normalization of interstellar FUV flux at $1000\ \AA$:
$$\U\equiv J_{1000\AA}/J_{\rm MW},$$ used throughout this paper, is
also defined to be in the units of the typical Milky Way value $J_{\rm
MW}=10^6\dim{photons}\,\dim{cm}^{-2}\,\dim{s}^{-1}\,\dim{ster}^{-1}\,\dim{eV}^{-1}$
\citep{ism:d78,ism:mmp83}. We explore the range of $\U$ from $0.1$ to
$100$ in our test simulations.

The star formation model in our simulations closely follows the
recipe 2 of \citet{ng:gtk09} with small numerical modifications.
Namely, the rate of star formation in each computational cell with
molecular fraction $f_{\rm H_2}\geq 0.1$ is evaluated as
\begin{equation}
  \frac{d\rho_\star}{dt} = \epsilon_{\rm
  SF}\frac{\rho_\H2}{\tau_{\rm SF}},
  \label{eq:sf}
\end{equation}
where the time scale for star formation is defined as $\tau_{\rm SF} =
\min(\tau_{\rm ff},\tau_{\max})$. We follows the definition of
\citet{sfr:kt07} for the gas free-fall time,
\[
  \tau_{\rm ff} = \sqrt{\frac{3\pi}{32G\rho}}
\]
(here $\rho$ is the total mass density, including helium), and
$\tau_{\max}$ is the free-fall time in the gas with
$n_{\rm SF}=50\dim{cm}^{-3}$. We adopt $\epsilon_{\rm SF} = 0.005$, which
is lower than the value we adopted in \citet{ng:gtk09} and is still
within the range of values advocated by \citet{sfr:kt07}. The lower
value of $\epsilon_{\rm SF}$ that we adopt provides a better fit the
THINGS measurements of the KS relation \citep{sfr:blwb08}.

The $\tau_{\rm sf}$ we adopt assumes that in low density cells, in
which molecular fraction $f_{\H2}$ is below unity, star formation proceeds
mainly in unresolved molecular clouds on subgrid scales. This
assumption then also motivates setting the maximum free fall time to
$\tau_{\rm max}$ corresponding to the number density of $50\,\,\rm
cm^{-3}$ typical average density of molecular clouds.  The
$f_{\H2}<1$ in these cells then can be viewed as reflecting the
fraction of the total gas in such star forming molecular clouds, which
themselves have $f_{\H2}=1$, rather than incomplete conversion of the
atomic gas into the molecular form inside the clouds.

As we show below (see Fig.~\ref{fig:sflpars1} and discussion in
\S~\ref{sec:sfl}), the KS relation in our simulations is not very
sensitive to variations of $\epsilon_{\rm SF}$ between $0.005$ and
$0.01$ and $n_{\rm SF}$ between $10$ and $50\,\,\rm cm^{-3}$.

%--------------------------------------------------------------
\section{The atomic-to-molecular gas transition}
\label{sec:fh2}
%--------------------------------------------------------------

\begin{figure}[t]
\includegraphics[scale=0.43]{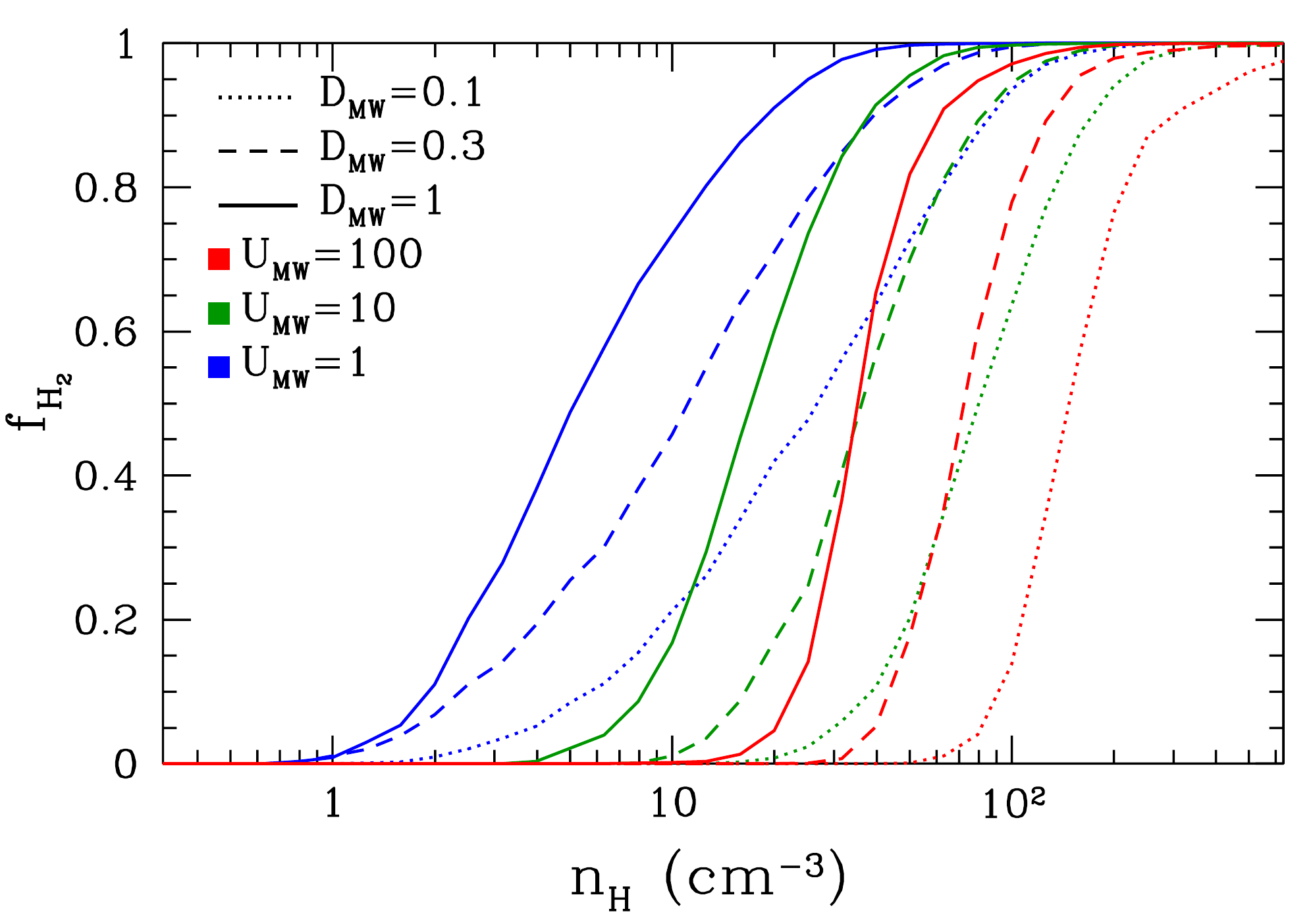}
%\epsscale{1.15} \plotone{\figname{fh2den1.ps}}
\caption{Average atomic-to-molecular gas transition as a function
  of total hydrogen number density for 9 test simulations (as
  distinguished by colors and line styles).} 
\label{fig:fh2den}
\end{figure}

The effect of two primary parameters, the dust-to-gas ratio $\D$ and
the interstellar FUV flux $\U$, on the transition from atomic
to molecular gas is illustrated in Figure \ref{fig:fh2den} as a function
of the total hydrogen density, $n_\Ht \equiv n_\HI + n_\HII + 2n_\H2$
(the contribution of ionized gas $n_\HII$ is negligible for
densities shown in Figure \ref{fig:fh2den}). As can be seen from the
figure, both parameters 
affect the atomic-to-molecular transition in a non-trivial way.

This scaling can be understood approximately if we ignore all physical
processes except the formation of molecular hydrogen on dust and
dissociation of molecular hydrogen by the UV radiation in the
Lyman-Werner band. This is necessarily an approximation, as many other
processes are indeed important for the detailed balance of molecular
hydrogen (see Appendix), but the formation on dust and
photo-dissociation are the dominant processes that control the
atomic-to-molecular gas transition under normal ISM conditions. In
this approximation, the equilibrium abundance of molecular hydrogen
can be determined from the balance of the formation and dissociation
rates (cf. Appendix) 
\begin{equation}
  n_\H2 \Gamma_{\rm LW} S_\H2(N_\H2) e^{-\sigma_{\rm LW} N_\Ht} = R_D
  n_\Ht n_\HI,
  \label{eq:h2bal}
\end{equation}
where $\Gamma_{\rm LW}=\U\Gamma_0$ is the free space photo-destruction
rate and $R_D$ and $\sigma_{\rm LW}$ are given by Equation
(\ref{eq:pardefs}). The atomic gas becomes molecular only due to
self-shielding and shielding by dust (the last two factors on the
left-hand-side of Equation (\ref{eq:h2bal})). If the FUV flux is not
too strong, the self-shielding by molecular hydrogen dominates; in
this limit dust absorption can be neglected and Equation
(\ref{eq:h2bal}) becomes
\[
  \frac{f_\H2}{1-f_\H2} = \frac{\D}{\U} n_\Ht \frac{R_0}{\Gamma_0
  S_{\H2}},
\]
where $f_\H2 \equiv n_\H2/n_\Ht$ and we ignore ionized gas. For our
ansatz for the self-shielding factor $S_\H2 \propto n_\H2^{-3/4}$
(Equation (\ref{eq:sh2})), so that
\[
  \frac{f_\H2^{1/4}}{1-f_\H2} \propto \frac{\D}{\U} n_\Ht^{7/4}.
\]
Thus, the characteristic density at which molecular hydrogen fraction
reaches a particular value (e.g., 50\%) scales
with the dust-to-gas ratio $\D$ and the FUX radiation flux
$\U$ as
\begin{equation}
  n_\Ht \propto \left(\frac{\U}{\D}\right)^{4/7}.
  \label{eq:h2scale1}
\end{equation}

In the opposite regime of large $\U$, the shielding by dust is
expected to dominate over self-shielding, because self-shielding is a
gradual function of the gas column density and may not be able to
provide the required shielding for sufficiently large UV fluxes. In
this regime, Equation (\ref{eq:h2bal}) becomes
\[
  \frac{f_\H2}{1-f_\H2} = \frac{\D}{\U} n_\Ht
  \frac{R_0}{\Gamma_0 S_\H2} e^{\D \sigma_0 N_\Ht}
\]
and the exponential factor is now large, so the characteristic
\emph{column} density for the atomic-to-molecular transition is 
\begin{equation}
  N_\Ht \propto \frac{\ln(\U/\D)}{\D}.
  \label{eq:h2scale2}
\end{equation}
Thus, as \citet{ng:gtk09} mention, in the regime where dust shielding
dominates, the dependence of the characteristic column density on the
FUV flux $\U$ is only logarithmic.

There is no way to convert between the characteristic column density
and the physical gas density easily. Nevertheless, the following
simple fitting formula captures the average dependence of the
atomic-to-molecular transition on the dust-to-gas ratio and the
FUV flux in our simulations:
\begin{equation}
  f_\H2 \approx \frac{1}{1+\exp\left(-4x-3x^3\right)},
  \label{eq:fh2fit}
\end{equation}
where $x$ is given by
\begin{equation}
  x \equiv \Lambda^{3/7} \ln\left(\D\frac{n_\Ht}{\Lambda n_{\ast}}\right).
  \label{eq:xdef}
\end{equation}
Here $n_{\ast}=25\dim{cm}^{-3}$, $\Lambda$ is 
\begin{equation}
  \Lambda \equiv \ln\left(1+g\D^{3/7}\left(\U/15\right)^{4/7}\right), 
  \label{eq:lamdef}
\end{equation}
and $g$ is a fudge factor to approximately account for the transition between
the two regimes: $g\approx 1$ when self-shielding dominates and
$g \propto \D^{-1}$ when dust shielding dominates. 

\begin{figure}[t]
\includegraphics[scale=0.43]{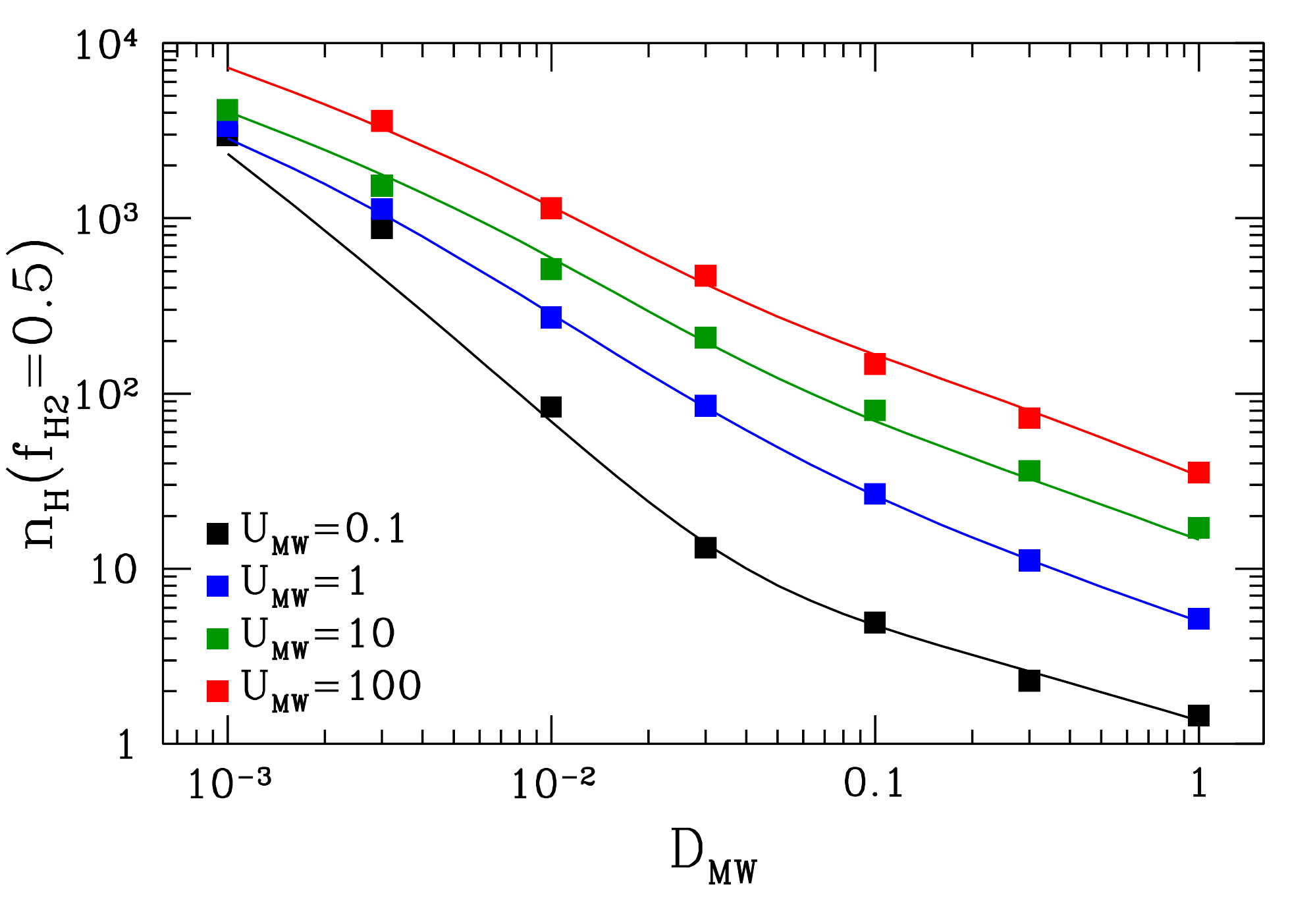}
%\epsscale{1.15}
%\plotone{\figname{nthd2g.ps}}
\caption{Average total hydrogen number density of
  atomic-to-molecular gas transition (defined as $f_\H2=0.5$) as a
  function of the scaled dust-to-gas ratio $\D$ and the FUV flux $\U$
  for all our test simulations. The point $(\D=0.001,\U=100)$ is
  missing because the resolution of our simulations is insufficient to
  capture the atomic-to-molecular transition in such extreme
  conditions. Solid lines show fitting formula of
  Equation (\ref{eq:nthfit}).\newline}
\label{fig:nthfit}
\end{figure}

We adopt the following fitting formula for the quantity $g$:
\[
  g = \frac{1 + \alpha s + s^2}{1 + s}
\]
where 
\[
  s \equiv \frac{0.04}{D_{\ast}+\D}, \ \ \ \alpha=5\frac{\U/2}{1+(\U/2)^2},
\]
and
\[
  D_{\ast} = 1.5\times 10^{-3}\times\ln\left(1+(3\U)^{1.7}\right)
\]
describes the transition to the regime when formation of $\H2$ via
the gas phase reactions dominates. 

Figure \ref{fig:nthfit} shows the value of the total (molecular,
atomic, and ionized - although the contribution of ionized gas in all
equations in this section is completely negligible) hydrogen density
at which molecular fraction reaches $f_\H2=0.5$ ($x=0$). Our fitting
formulae give the following approximate expression for this density:
\begin{equation}
  n_\AM \equiv n_\Ht(f_\H2=0.5)\approx n_{\ast} \frac{\Lambda}{\D}.
  \label{eq:nthfit}
\end{equation}
This equation is a better approximation than the the simple
step-function ansatz proposed in \citep{ng:gtk09}. Figure
\ref{fig:nthfit} demonstrates that Equation \ref{eq:nthfit} indeed
provides an accurate model for the dependence of $n_\Ht(f_\H2=0.5)$ on
$\D$ and $\U$.

\begin{figure}[t]
\includegraphics[scale=0.43]{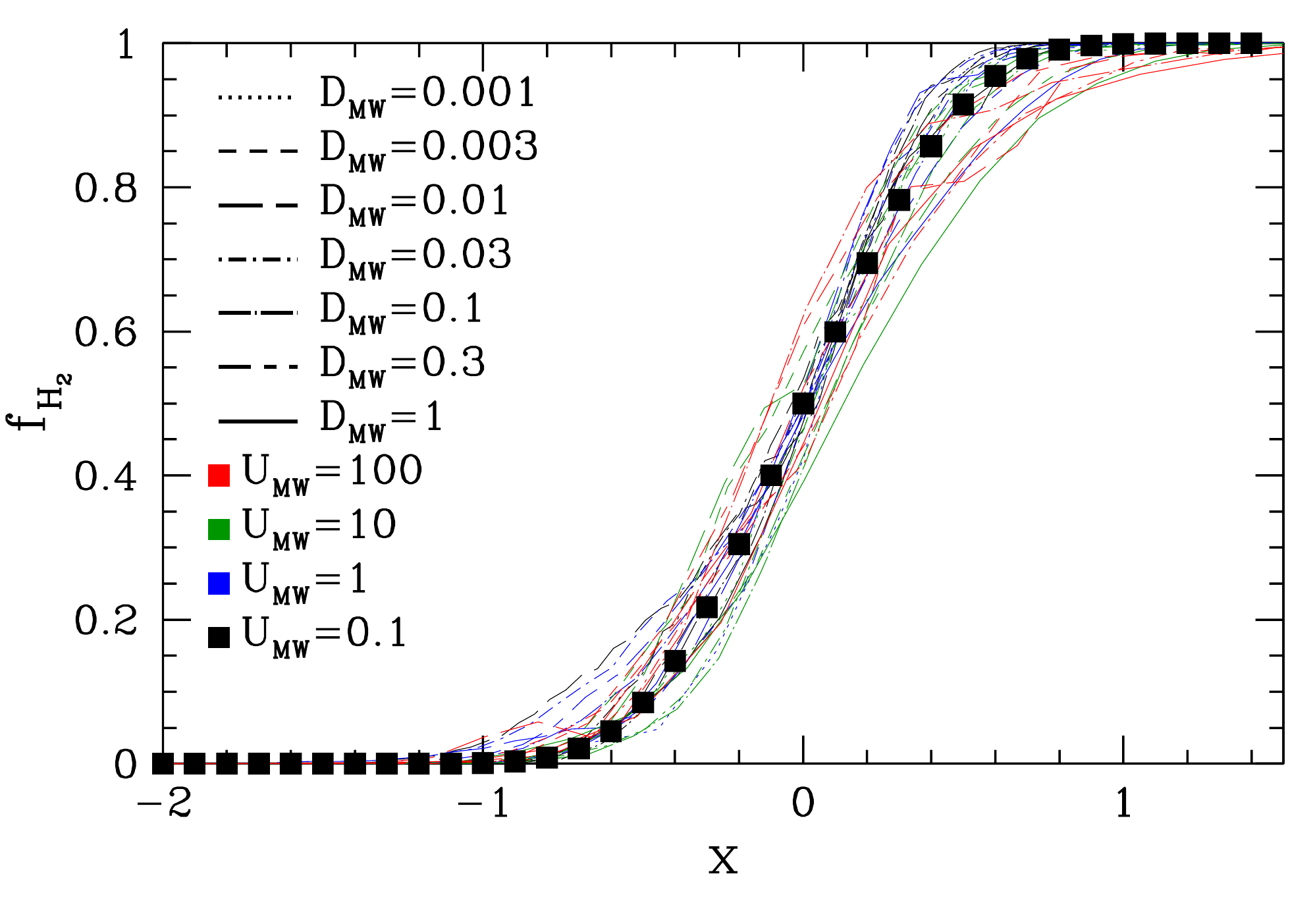}
\includegraphics[scale=0.43]{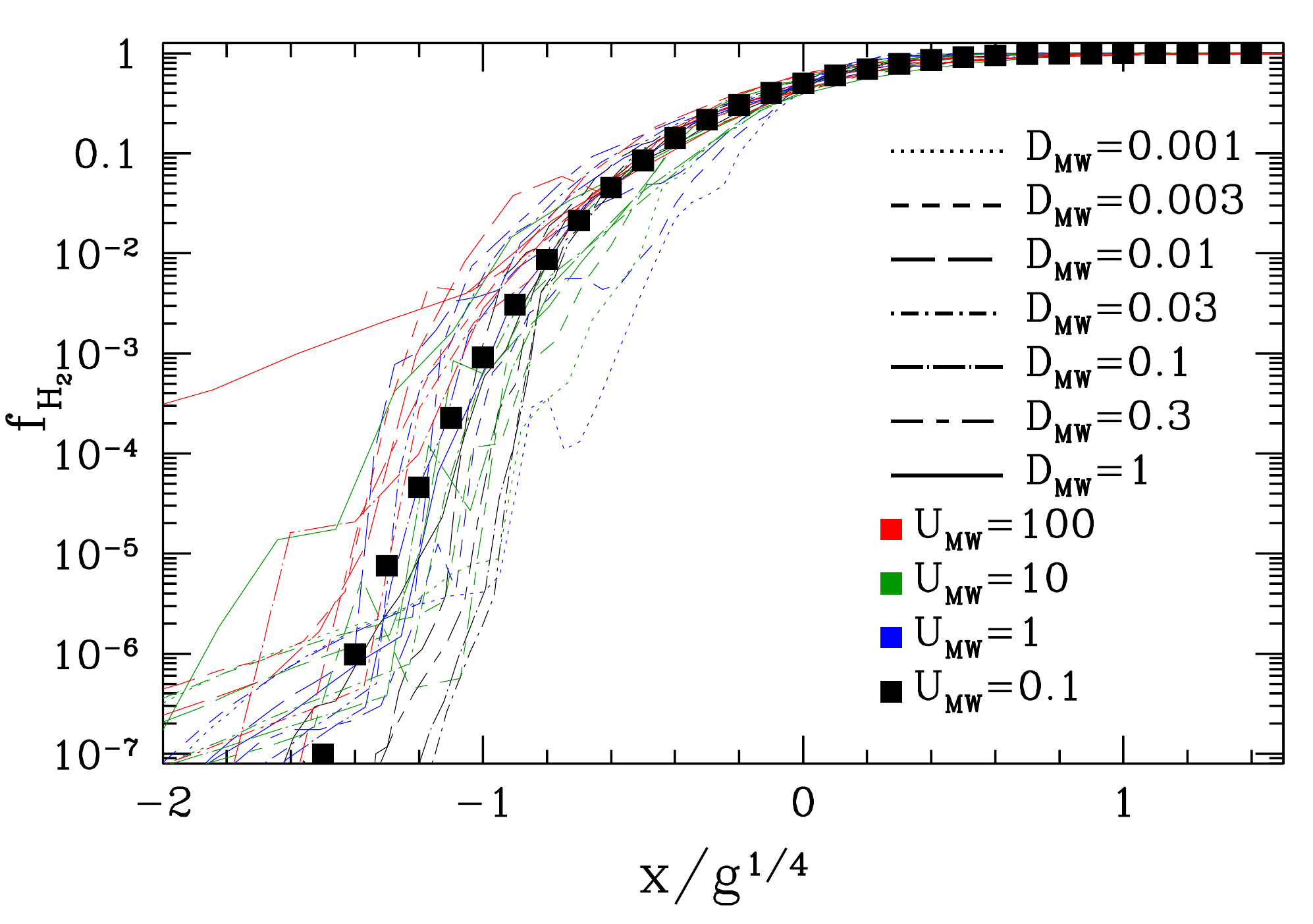}
%\epsscale{1.15}
%\plotone{\figname{fh2den2a.ps}}
%\plotone{\figname{fh2den2b.ps}}
\caption{Average atomic-to-molecular gas transition as a function
  of the factorized variable $x$ (Equation (\ref{eq:xdef})) for all our
  test simulations (as distinguished by colors and line styles). The
  top panel shows the linear scaling of the $y$ axis (most relevant
  for modeling star formation) while the bottom panel shows the $y$
  axis in log. Black squares on the right panel show the approximation
  from Equation (\ref{eq:fh2fit}).}
\label{fig:fh2fit}
\end{figure}

Figure \ref{fig:fh2fit} shows that Equation (\ref{eq:fh2fit}) works
well for $f_\H2\gtrsim 0.1$ for all simulated cases (4 values of $\U$
and 7 values of $\D$), but it becomes somewhat less accurate for lower
molecular fractions.  The accuracy in the low $f_\H2$ regime can be
improved with a simple modification: replacing $x$ in Equation
(\ref{eq:fh2fit}) with $x/g^{1/4}$. This change provides a more
accurate fit for the range $10^{-5}\lesssim f_\H2\lesssim 0.1$, but is
less accurate than the above approximation for $f_\H2>0.1$. Given that
for modeling star formation the range $f_\H2\gtrsim 0.1$ is most
relevant, we use the unmodified form of our fit as the fiducial
approximation.

Neither form of this fit describes the equilibrium $\H2$ abundance
($f_\H2 \sim 10^{-6} - 10^{-8}$) in the Warm Interstellar Medium. Such
a small abundance is, of course, not relevant to star formation.

%----------------------------------------------
\section{The Kennicutt-Schmidt relation and its dependence on the dust-to-gas ratio and the FUV flux}
\label{sec:sfl}
%----------------------------------------------

The physics of the transition from atomic to molecular phase,
discussed in the previous section, controls which local regions within
the interstellar medium of simulated galaxies have high-molecular
fraction and, hence, become the sites of star formation. Although the
local rate of star formation in these regions is sensitive to the
parameters of the $\H2$ formation model and star formation recipe, the
global star formation rate surface density on larger, kiloparsec
scales depends on the density and UV flux distribution within larger
scales that are modeled self-consistently in the
simulations. Therefore, once we fix the parameters of the model
controlling the chemistry and star formation on small scales, we can
examine the {\it predicted} KS relation between the surface densities
of various gas phases and the surface density of star formation
averaged on large scale.

Observationally, only the surface densities of atomic and molecular
gas are directly measured and included in the estimate of the ``total''
surface gas density, $\Sgas$. However, as we demonstrate
below, the ionized gas may contribute significantly to the total gas
surface density under some conditions. Therefore, we deliberately
avoid using the ambiguous notation $\Sgas$ and instead use the following 
notation explicitly indicating the components that are included in 
the surface density:
\[
  \Stot \equiv \Shii + \Shi + \Smol,
\]
for the total surface density, uncluding both neutral and ionized gas, and 
\[
  \Sntr \equiv \Shi + \Smol,
\]
for the surface density, including only neutral atomic and molecular
gas. Note that we follow the observational practice and do not include
contribution of helium in the above gas surface densities.  We
emphasize again that in observational work the total gas density is
commonly identified with this second quantity, $\Sgas = \Sntr$.

As we mentioned in the previous section, this distinction is
unnecessary for studying the atomic-to-molecular gas transition on
small scales, because the fraction of ionized gas is always small at
densities at which the molecular fraction is significant. In other
words, high-$f_\H2$ regions are always surrounded by neutral atomic
envelopes containing little ionized gas. However, regions of a
kiloparsec scale can contain a mix of different ISM phases: from
low-density ionized gas to high-density, molecular regions. In fact,
diffuse ionized ISM gas is ubiquitous in nearby galaxies
\citep[e.g.,][]{hoopes_etal03}. The warm ($\sim 10^4$~K) diffuse
ionized gas is present both inside the disk and at large distances (up
to $\sim 2-4\dim{kpc}$) from the midplane both in the Milky Way
\citep{reynolds89,reynolds91,gaensler_etal08} and other nearby
galaxies \citep[e.g.,][see \citeauthor{haffner_etal09}
\citeyear{haffner_etal09} for
review]{hoopes_etal99,collins_etal00,rossa_dettmar03}. This ionized
gas can be a significant fraction of the total gas density.  In the
Milky Way, for example, the warm ionized gas accounts for $\sim 25\%$
of the total hydrogen column density of the disk
\citep{reynolds91,haffner_etal09}. One has to keep in mind the
possible presence of such gas in theoretical interpretations of the KS
relation.

\begin{figure}[t]
\includegraphics[scale=0.43]{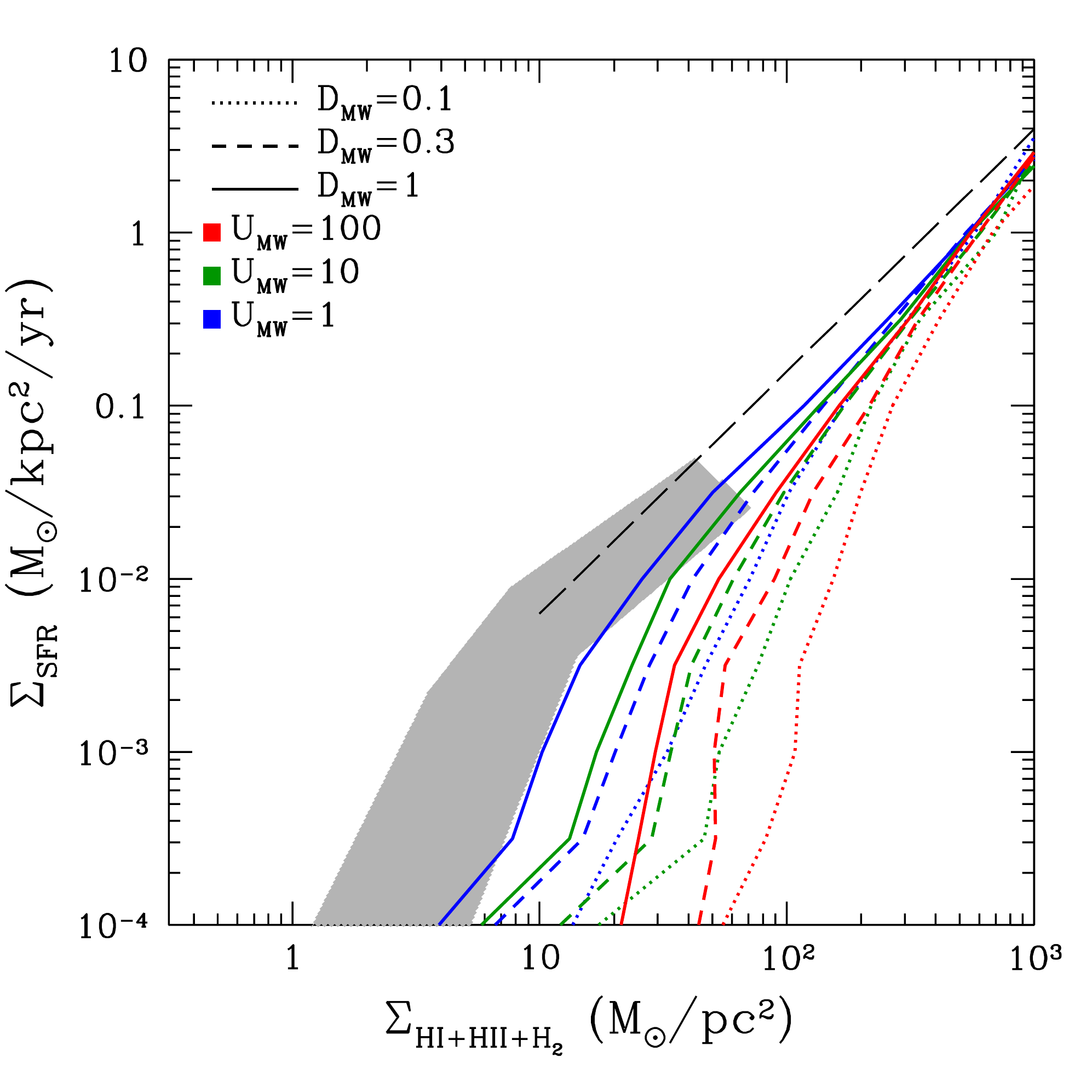}
%\epsscale{1.15}
%\plotone{\figname{sflaw3.ps}}
\caption{Relation between $\Ssfr$ and the \emph{total} surface
  density of gas (atomic, molecular, and ionized) for 9 different
  representative combinations of dust-to-gas ratio and the
  interstellar FUV flux (colored lines). The long-dashed line is the
  best fit relation of \citet{sfr:k98a} for $z\approx 0$ galaxies. The
  gray shaded area shows the KS relation for the local dwarf and
  normal spiral galaxies measured by the THINGS project
  \citep{sfr:blwb08}.}
\label{fig:sfltot}
\end{figure}

For comparison with observations, the star formation rate in the
simulations is averaged over $20\dim{Myr}$ and the gas and SFR surface
densities are averaged on the scale of $500\dim{pc}$. This specific
choice corresponds to the averaging spatial scale and star formation
indicator used in the THINGS measurements
\citep{sfr:srcb07,sfr:blwb08}. We tested the sensitivity of the
predicted KS relation to the specific choice of the averaging temporal
and spatial scales; such a comparison is presented in the Appendix
(see Figure~\ref{fig:avgsfl}). Overall, the KS relation is robust to
changes of spatial and temporal averaging scales with the range
$0.5-2.0\dim{kpc}$ and $20-100\dim{Myr}$, respectively.  Some modest trends are
observed, but these are in general agreement with observations.

In Figure \ref{fig:sfltot} we show the relation between $\Ssfr$ and
the \emph{total} surface density of gas (atomic, molecular, and
ionized), $\Stot$, for nine different representative combinations of
dust-to-gas ratio and the interstellar FUV flux $\D$ and
$\U$.  As could be expected, both the dust-to-gas ratio $\D$ and the
UV flux $\U$ affect the relation significantly by affecting the
atomic-to-molecular transition and the fraction of neutral gas in the
ISM patches. Notably, the predicted $\Ssfr-\Stot$ relation does not
agree with observations for any combination of $\U$ and $\D$.

\begin{figure*}[t]
\includegraphics[scale=0.43]{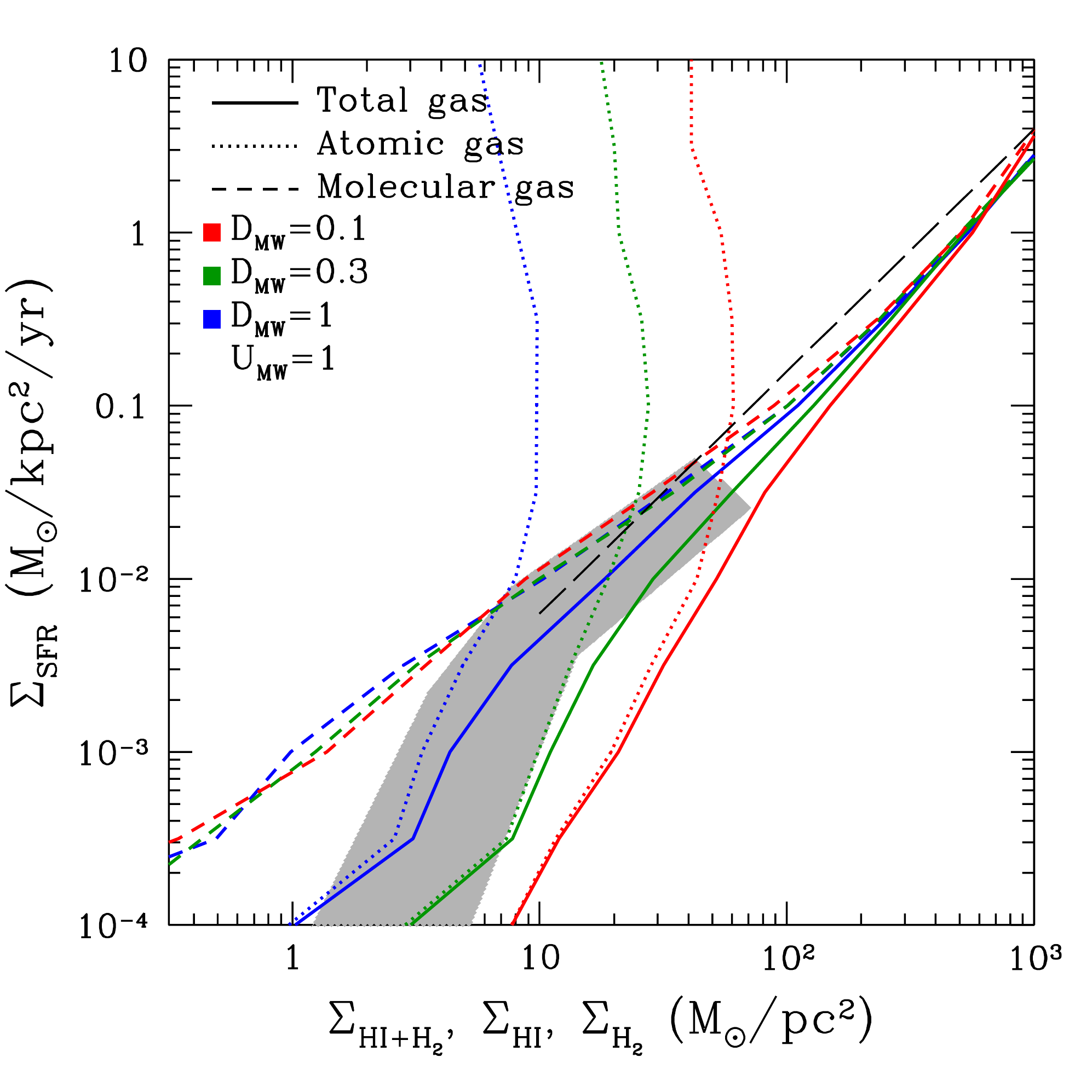}\hspace{1cm}
\includegraphics[scale=0.43]{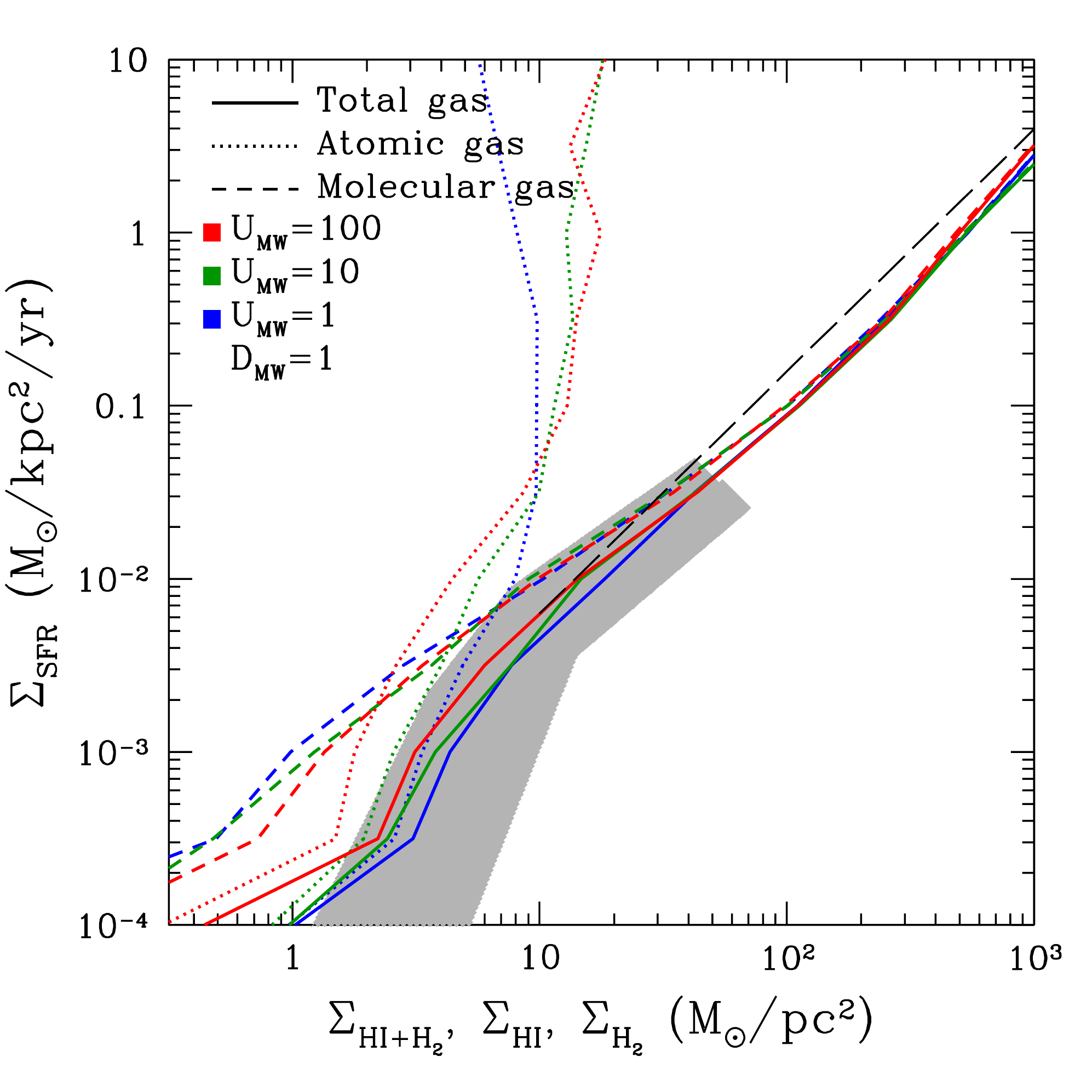}
%\epsscale{1.15}
%\plottwo{\figname{sflaw4a.ps}}{\figname{sflaw4b.ps}}
\caption{KS relations for the \emph{neutral} gas (atomic and
  molecular) predicted in models with different representative values
  for the dust-to-gas ratio (left panel) and the interstellar FUV
  radiation fluxe (right panel) are shown as colored lines. Dotted,
  short-dashed, and solid lines show the relation between $\Ssfr$ and
  $\Shi$, $\Smol$, and $\Sntr$ individually.  The observed relations
  (long-dashed line and gray band) are the same as in Fig.\
  \ref{fig:sfltot}.}
\label{fig:sflntr}
\end{figure*}

However, as we emphasized above, observational measurements often do
not account for the contribution of ionized gas to surface density. We
therefore present a separate prediction for the KS relation for the
neutral gas only in Figure \ref{fig:sflntr} for a representative
subset of our test simulations. This figure demonstrates that the
predicted $\Ssfr-\Sntr$ relation for the parameter values
representative of local galaxies ($\D\sim1$ and any value of $\U$) is
in good agreement with both the older measurement of \citet{sfr:k98a}
and with the recent measurements by The $\HI$ Nearby Galaxy Survey
(THINGS) \citep{sfr:blwb08}. In particular, our model approximately
reproduces the rapid decrease of the SFR and increase of the scatter
at $\Sntr < 10 \Msun\dim{pc}^{-2}$ and the change in the slope of the
star formation rate vs gas surface density from $\Ssfr \propto
\Sntr^{1.0}$ to $\Ssfr \propto \Sntr^{1.4}$ at $\Sntr\approx 10^2
\Msun\dim{pc}^{-2}$.

The KS relations shown in Fig.~\ref{fig:sfltot} and \ref{fig:sflntr}
can be accurately described by a simple fitting
formula. Since stars only form in molecular gas, the star formation
rate surface density is proportional to the surface density of
molecular gas,
\[
  \Ssfr = \frac{1}{\tau_{\rm SF}}\Smol,
\]
where $\tau_{\rm SF}$ is the time scale for star formation (that may
itself depend on the molecular gas surface density). If the neutral
gas surface density $\Sntr$ is used as an argument, the reduced
star formation rate at low gas surface density needs to be taken
into account,
\begin{equation}
  \Ssfr = \frac{1}{\tau_{\rm SF}}\frac{\Sntr}
  {\left(1+\Sigma_{\ast}/\Sntr\right)^2},
  \label{eq:ksfit}
\end{equation}
where $\Sigma_{\ast}$ is the characteristic surface density of neutral
gas at which the relation steepens. At large gas surface densities
(i.e., $\Sigma_{\ast}\ll \Sntr$) we have:
\begin{equation}
  \Ssfr \approx \frac{1}{\tau_{\rm SF}} \left(\Sntr-\Shi^\infty\right),
  \label{eq:ksinf}
\end{equation}
where $\Shi^\infty$ is the saturation value of $\HI$ surface density, 
i.e.\ the maximum $\Shi$ reached by gas as its total surface density
increases to large values.  Note that comparison of this equation with
the formula of Equation \ref{eq:ksfit} shows that
\[
  \Sigma_{\ast} = \frac{\Shi^\infty}{2}.
\]

\begin{figure}[t]
\includegraphics[scale=0.43]{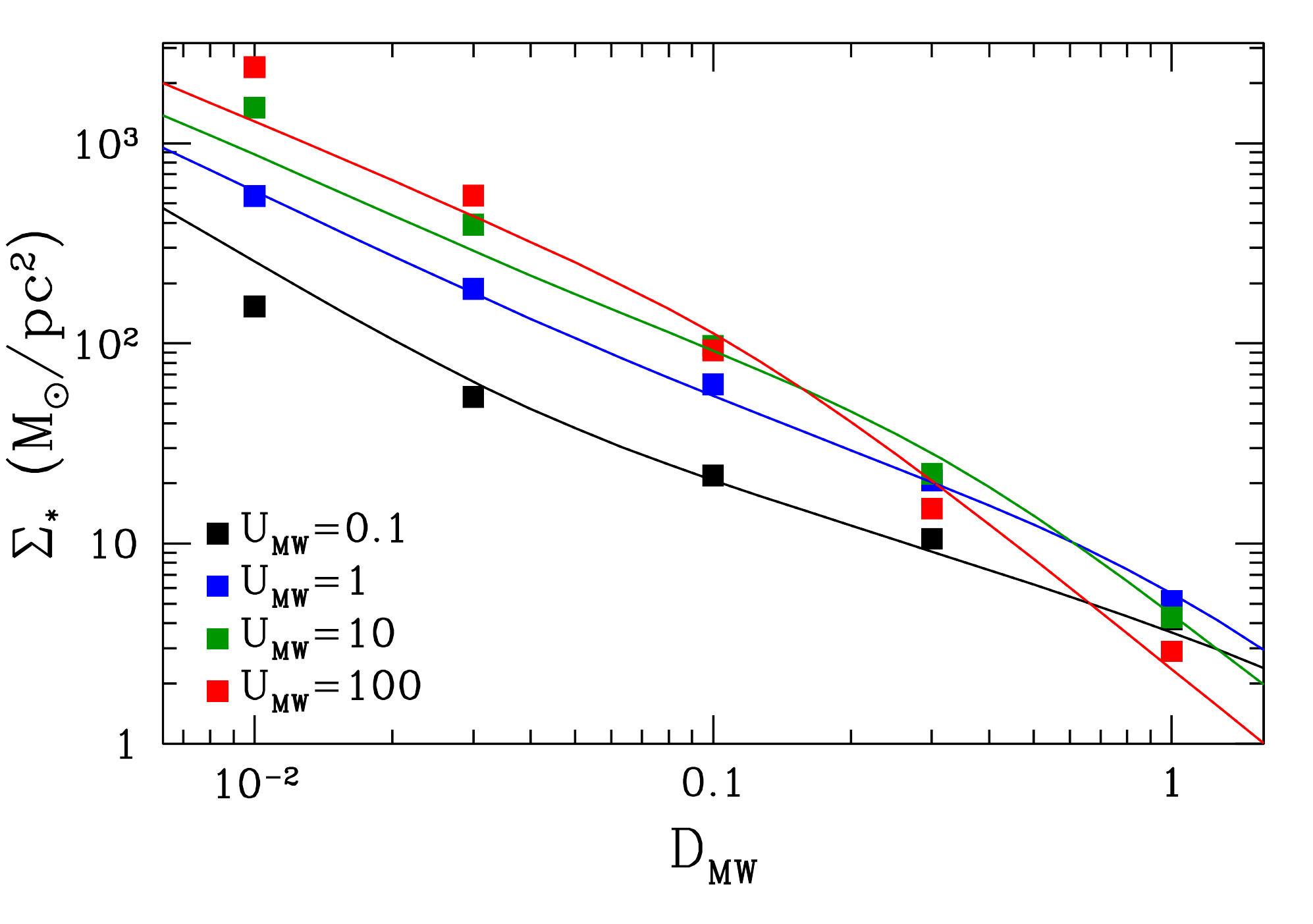}
%\epsscale{1.15}
%\plotone{\figname{sthd2g.ps}}
\caption{Characteristic threshold surface density
  $\Sigma_{\ast}$ as a function of two main
  parameters $\D$ and $\U$ for all our test simulations. Cases with
  $\D<0.01$ are not shown, as in our simulations gas at such low values
  of the dust-to-gas ratio never becomes fully molecular on
  $500\dim{pc}$ scale (and, thus, $\tilde{\Sigma}_{\rm SFR}$ cannot be
  determined). The solid lines show the fitting formula of
  Equation (\ref{eq:sstar}).\newline}
\label{fig:sstar}
\end{figure}

Figure \ref{fig:sflntr} demonstrates that, while the dust-to-gas ratio $\D$
plays the dominant role in controlling the turnover in the
$\Ssfr-\Sntr$ relation at low surface densities for $\D\gtrsim0.1$,
this is no longer the case at lower dust-to-gas ratios. Figure
\ref{fig:sstar} shows the dependence of the characteristic
``threshold'' surface density $\Sigma_{\ast}$ on $\U$ and $\D$ for the
full suite of our models. At $\D\lesssim 0.1$, $\Sigma_{\ast}$ changes
by an order of magnitude for $\U$ changing by three orders of magnitude
between $0.1$ and $100$. Thus, although dependence of the KS relation
on the FUV flux for higher dust content systems is expected to be
weak, it can be stronger for dwarf galaxies at $z\approx 0$ and in
high-$z$ galaxies with low dust-to-gas ratios.

The dependence of the $\HI$ saturation surface density on our two main
parameters can be understood qualitatively if we assume that the
density distribution in the ISM is approximately self-similar. Let us
consider a large-scale region over which we measure the total hydrogen
surface density $\Stot$. Within this region the total hydrogen density
has some density probability function (defined as a fraction of
surface density contributed by gas of a given density $n_\Ht$), which
in general depends on $\Stot$: $\phi(n_\Ht,\Stot)$. If the density
distribution is self-similar a region with a higher surface density
will have more dense gas, i.e.
\[
  \phi(n_\Ht,\Stot) = \psi(\xi),
\]
where $\xi = n_\Ht/\Stot$ and 
\[
  \int_0^\infty \psi(\xi) d\xi = 1.
\]
The atomic hydrogen surface density is then simply
\[
  \Shi = \int_0^\infty f_\HI \phi dn = \Stot \int_0^\infty f_\HI
  \psi(\xi) d\xi.
\]

If we assume that most of the atomic hydrogen mass is at densities
near the atomic-to-molecular transition density $n_\AM$ (which is the
case in our simulations), then we can use our parametrization from
Equation (\ref{eq:fh2fit}) a function of factorized variable $x$
(Equation (\ref{eq:xdef})), so that
\[
  \frac{d\xi}{dx} = \frac{n_\Ht}{\Stot} \frac{dx}{\Lambda^{3/7}},
\]
and 
\[
  \Shi = \frac{1}{\Lambda^{3/7}} \int_{-\infty}^\infty f_\HI n_\Ht
  \psi(\xi) dx.
\]
The last integral cannot be taken exactly, but given that the
atomic-to-molecular transition is a rather steep function of the
gas density, the integral can be approximated as
\begin{eqnarray}
  \Shi & \approx & \frac{1}{\Lambda^{3/7}} \left.\left(f_\HI n_\Ht
  \psi\right)\right|_{\AM} {\Delta x} \nonumber \\
  & = & \frac{1}{\Lambda^{3/7}} \frac{1}{2} n_{\AM} \psi(\xi_\AM/2) {\Delta x},
  \label{eq:shi}
\end{eqnarray}
where $\Delta x \sim 1$ is the width of the atomic-to-molecular
transition ($f_\HI=f_\H2=0.5$) in the variable $x$, which should be essentially
independent of any physical parameter.

The saturation $\HI$ surface density $\Shi^\infty$ is obtained from
Equation (\ref{eq:shi}) in the limit of $\Stot \rightarrow \infty$, in
which case the argument of $\psi$ in Equation (\ref{eq:shi}) can be
replaced with zero, and we finally obtain
\begin{equation}
  \Shi^\infty \approx  \psi(0) \frac{n_\AM}{2\Lambda^{3/7}} {\Delta
  x} \propto  \frac{\Lambda^{4/7}}{\D}.
  \label{eq:shisat}
\end{equation}
We find that this scaling works well in our simulations, except in the
limit of large $\D$ and large $\U$,  when the density of 
the ionized-to-atomic transition is not negligible
compared to the density of the atomic-to-molecular transition. As
a consequence, the contribution of the ionized gas is not negligible
compared to the atomic gas, which leads to a decrease of
$\Shi^\infty$ compared to the value predicted by Equation
(\ref{eq:shisat}). In the extreme case we consider ($\D=1$, $\U=100$)
the saturation $\HII$ surface density is 3-4 times higher than the
saturation $\HI$ surface density.

\begin{figure}[t]
\includegraphics[scale=0.43]{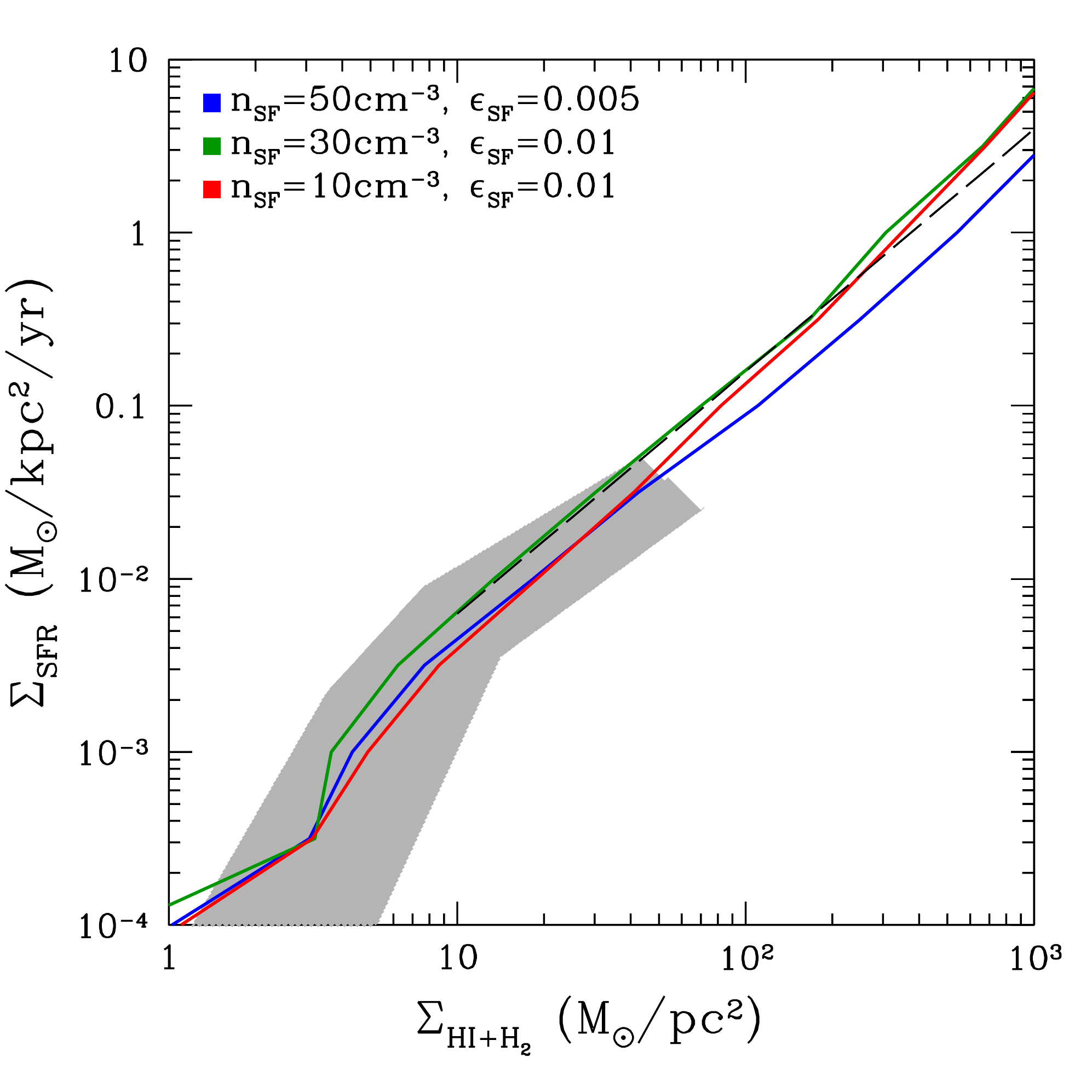}
%\epsscale{1.15}
%\plotone{\figname{sflaw7.ps}}
\caption{Dependence of the KS relation for the neutral gas
  (atomic and molecular) on the parameters of the star formation
  recipe (\ref{eq:sf}). The long-dashed line is the best fit relation
  of \citet{sfr:k98a} for $z\approx 0$ galaxies. The gray shaded area
  shows the KS relation for the local dwarf and normal spiral galaxies
  measured by the THINGS project \citep{sfr:blwb08}. }
\label{fig:sflpars1}
\end{figure}

The following simple fitting formula corrects for this deficiency and
provides a good fit for the characteristic ``threshold'' surface
density, $\Sigma_{\ast}$, and HI saturation surface density
$\Shi^\infty\equiv 2\Sigma_{\ast}$ in all test cases we consider,
 \begin{equation}
  \Sigma_{\ast} = 20\Msun\dim{pc}^{-2}
  \frac{\Lambda^{4/7}}{\D} \frac{1}{\sqrt{1+\U\D^2}}.
  \label{eq:sstar}
\end{equation}
The accuracy of this fitting formula is demonstrated in Figures
\ref{fig:sstar} and \ref{fig:sfall}. For very low values of
$\D\lesssim0.01$ the fit is not very accurate. This is most likely
due to the limited volume of our simulations: at such low dust-to-gas
ratios the atomic-to-molecular transition shifts to extremely high gas
densities, $n_\Ht\sim10^3\dim{cm}^{-3}$, and our simulations lack
$500\dim{pc}$ sized regions that would be dominated by such dense
gas. Large volume simulations containing substantially more massive
galaxies will be need to test the accuracy of the fitting formula
(\ref{eq:sstar}) in this regime.

Finally, we have checked that our results are not particularly
sensitive to the specific choice of the fiducial parameters
$\epsilon_{\rm SF}$ and $n_{\rm SF}$. While the fiducial values
provide the best fit to the median values of THINGS
measurements \citep{sfr:blwb08}, a substantial variation in the adopted
values for these parameters has only mild effect on our results, as we
demonstrate in Figure \ref{fig:sflpars1}.

%--------------------------------------------------------
\section{Star formation recipes}
\subsection{Recipe for galaxy formation simulations}
\label{sec:SIMrecipe}
%--------------------------------------------------------

\begin{figure}[t]
\includegraphics[scale=0.43]{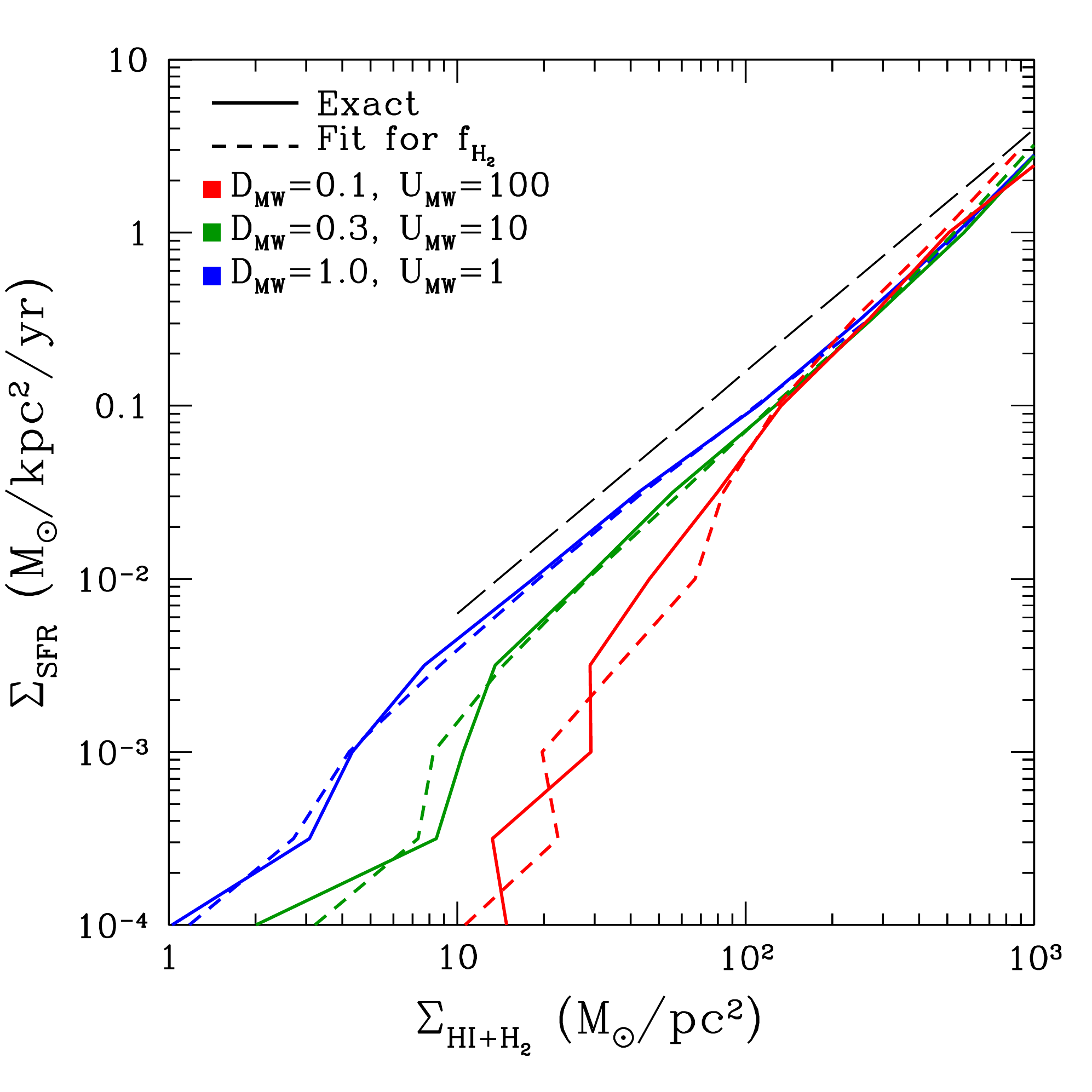}
%\epsscale{1.15}
%\plotone{\figname{sflaw6.ps}}
\caption{Comparison of the KS relation for the neutral gas (atomic and
  molecular) for the full simulations and test runs which used
  Equation (\ref{eq:fh2fit}) to estimate the molecular fraction in the
  gas for a representative subset of values for $\D$ and $\U$.}
\label{fig:sflpars2}
\end{figure}

In \S~\ref{sec:fh2} we have shown that atomic to molecular transition
density can be well fit by fitting functions as a function of
dust-to-gas ratio and FUV flux (e.g., Equation (\ref{eq:fh2fit})).  These
fitting functions are an approximation to the \emph{average}
dependence of the molecular fraction on the total hydrogen
density. The scatter in this relation around the mean may be important
for particular observational measurements of the molecular abundance
in the ISM. However, it is interesting to ask the question of whether
we can reproduce results of our full simulations by using the fit for
molecular fraction given by Equation (\ref{eq:fh2fit}) in star
formation recipe of Equation (\ref{eq:sf}), instead of the true $f_\H2$
calculated using our full chemistry model.  The results of such tests
are shown in Figure \ref{fig:sflpars2}, which demonstrates that using
the fit to $f_\H2(n_{\rm H})$ gives results closely matching results
of the full calculations.

This means that the approximation of Equation (\ref{eq:fh2fit}) can be
used to implement the $\H2$-based star formation recipe in galaxy
formation simulations that do not follow the full molecular chemistry,
provided that the resolution of the simulations is sufficiently high
($\sim100\dim{pc}$) and that the values for the parameters $\D$ and $\U$
could be estimated or assumed. The dust-to-gas ratio, $\D$ can be
estimated using local gas metallicity $Z$. 
Although the observed relation between $\D$ and $Z$ has a substantial scatter, on \emph{average} the dust-to-gas ratio appears to be directly proportional to the gas metallicity,
\[
  \D = \frac{Z}{Z_\odot},
\]
both for normal galaxies \citep{ism:i03,ism:ddbg07,ism:cpm08} and in low metallicity dwarfs \citep{ism:lf98,ism:h99,ism:cpm08,ism:m08}. Such a simple relation is, necessarily, a crude approximation, since not only the abundance, but even the properties of dust are known to be different in different galaxies.

Relating the local FUV flux $\U$ is trickier, but
sensible estimates can be made using the local SFR rate averaged on a
certain scale, as was done for example by
\citet{robertson_kravtsov08}. Given the steepness of the atomic to
molecular transition, the H$_2$-based star formation recipe amounts to
the metallicity and FUV flux dependent density threshold for star
formation.

%--------------------------------------------------------
\subsection{Star formation recipe for semi-analytic models}
\label{sec:SAMrecipe}
%--------------------------------------------------------

The dependence of the KS relation on the dust-to-gas ratio and the FUV flux in
our test simulations described in \S \ref{sec:sfl} can also be
encapsulated by a simple recipe. Such a recipe can be used in
semi-analytic models, in which radial dependence of gas surface
density, star formation, and chemical enrichment are modeled
explicitly
\citep[e.g.,][]{firmani_avilareese00,kravtsov_etal04,dutton_etal07}.

As we discussed above, the dependence of the KS relation on $\D$ and
$\U$ in our models is due to the dependence of the characteristic $\HI$ surface
density, $\Sigma_\ast$, on these variables. We therefore parameterize
the KS relation by the following fitting formula,
\begin{equation}
  \Ssfr = \frac{\tilde{\Sigma}_{\rm SFR}(\Sntr)}
  {\left(1+\Sigma_{\ast}/\Sntr\right)^2},
% I don't like -2 powers, they are easily to overlook
% \Ssfr = \tilde{\Sigma}_{\rm SFR}
% \left(1+\frac{\Sigma_{\ast}}{\Sntr}\right)^{-2},
  \label{eq:sfit}
\end{equation}
where $\Sigma_{\ast}$ is given by Equation (\ref{eq:sstar}) and
$\tilde{\Sigma}_{\rm SFR}(\Sntr)$ is the star formation rate in the fully
molecular gas at this surface density. For the latter, one can adopt
either the original Kennicutt fit \citep{sfr:k98a}:
\begin{equation}
  \tilde{\Sigma}_{\rm SFR, K} = 2.4\times10^{-4}
  \frac{\Msun}{\dim{kpc}^2\dim{yr}}
  \left(\frac{\Sntr}{1\Msun\dim{pc}^{-2}}\right)^{1.4}, 
  \label{eq:sfitk}
\end{equation}
or the fit suggested by the study of \citet{sfr:blwb08}:
\begin{equation}
  \tilde{\Sigma}_{\rm SFR, B} = \frac{\Sntr}{800\dim{Myr}}
  \max\left(1,\frac{\Sntr}{\Sigma_{\alpha}}\right)^{\alpha}. 
  \label{eq:sfitb}
\end{equation}
with the values of $\Sigma_{\alpha}\approx 200\Msun\dim{pc}^{-2}$ and $\alpha\approx 0.5$. 
Note that neither the slope at high surface densities $\alpha$ 
nor the characteristic surface density $\Sigma_{\alpha}$ at which the slope steepens are well constrained by the
current observations.

\begin{figure}[t]
\includegraphics[scale=0.43]{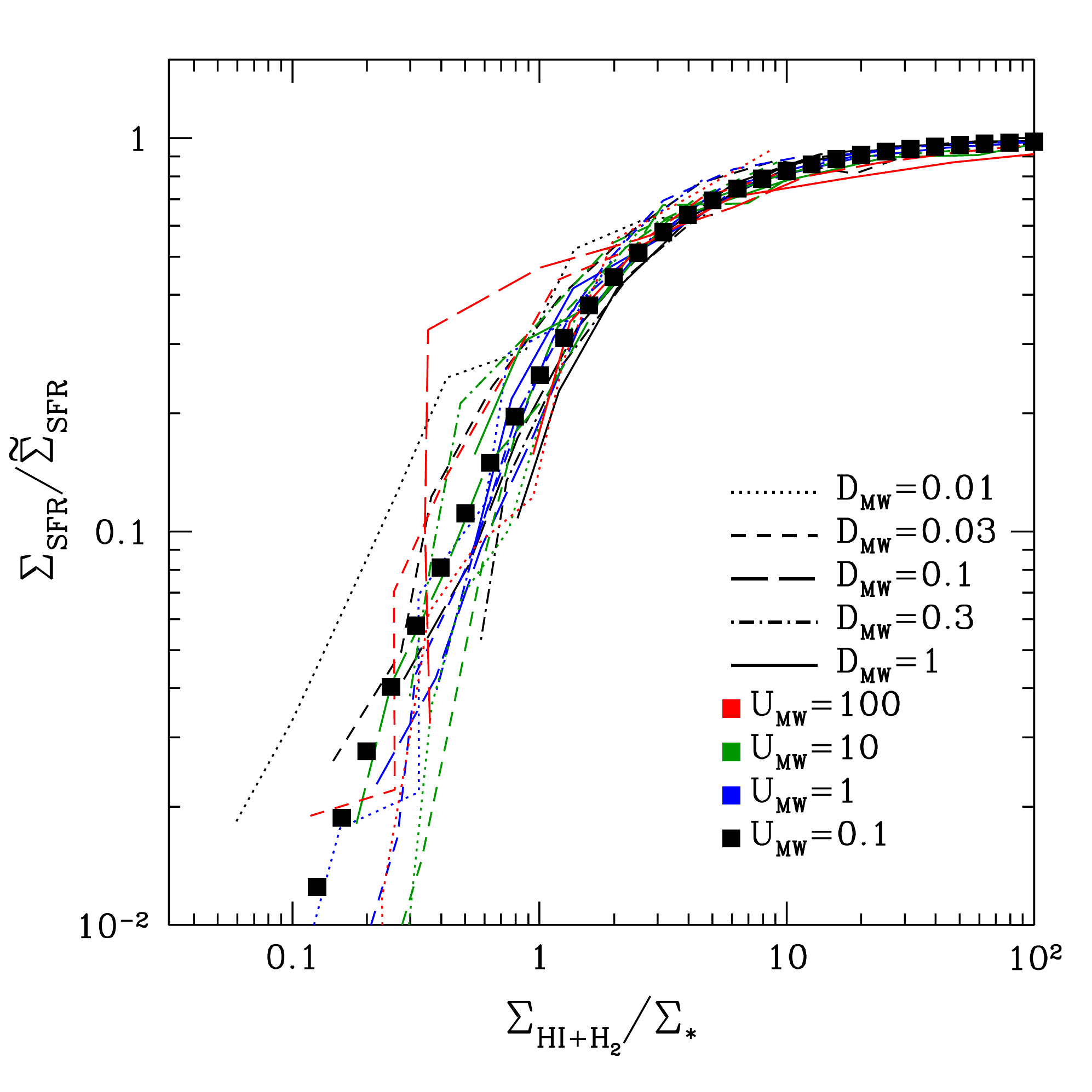}
%\epsscale{1.15}
%\plotone{\figname{sflaw5.ps}}
\caption{Scaled KS relation as a function of the neutral gas
  surface density, scaled by the characteristic surface density
  $\Sigma_{\ast}$. Black squares show the fitting formula (\ref{eq:sfit})
  with $\Sigma_{\ast}$ given by Equation (\ref{eq:sstar}) and
  $\tilde{\Sigma}_{\rm SFR}$ measured directly from the simulation as
  the star formation rate density in the molecular gas. Cases with
  $\D<0.01$ are not shown, as in our simulations gas at such low values
  of the dust-to-gas ratio never becomes fully molecular on
  $500\dim{pc}$ scale (and, thus, $\tilde{\Sigma}_{\rm SFR}$ cannot be
  measured).\label{fig:sfall}\newline}
\end{figure}

Figure \ref{fig:sfall} shows that the fitting formula for the KS
relation of Equation (\ref{eq:sfit}) together with Equation (\ref{eq:sstar})
reproduce the dependence of the KS relation on $\U$ and $\D$ in
simulations remarkably well. In semi-analytic models this formula can
be used if one has some prescription for estimating $\D$ and $\U$ in
model galaxies. As we noted in the previous sections, these
variables can be estimated approximately from the local metallicity of
the gas and local star formation rate.

%-----------------------------------
\section{Discussion and Conclusions}
\label{sec:discussion}
%-----------------------------------

We have presented results of a phenomenological model for formation of
molecular hydrogen and have illustrated the dependence of molecular
fraction on the gas density, dust-to-gas ratio, and far UV radiation
flux. We have also presented the large-scale Kennicutt-Schmidt
relation arising in our simulated galaxies when the local star
formation is based on the density of molecular (rather than total)
gas. Such approach allows us to avoid arbitrary density and
temperature thresholds typically used in star formation recipes. Our
results show that both the molecular fraction and the KS relation are
sensititive to the dust-to-gas ratio and the FUV flux,
although the sensitivity of the KS relation to the dust-to-gas ratio
is stronger than to the FUV flux.

We parameterize the dependencies observed in our simulations by
fitting formulae (\S~\ref{sec:fh2} and \ref{sec:sfl}), which can be
used to approximately account for $\H2$ formation and $\H2$-based star
formation in simulations, which do not include a full $\H2$ formation
model and radiative transfer (see \S~\ref{sec:SIMrecipe}). We
demonstrate that our fitting formulae, when applied to realistic
simulations, produce results that are close to those obtained in
simulations with the full $\H2$ formation model and radiative transfer
(Figure~\ref{fig:sflpars2}).

We also provide fitting formulae for the dust-to-gas and the FUV
radiation flux dependence of the KS relation that can be used in the
semi-analytic models of galaxy formation (\S~\ref{sec:SAMrecipe}). One
recent example of a model where such dependendcies can be relevant is
the study of \citet{dutton_etal10}.  The results of that study
indicate that the redshift evolution of SFR-$M_{\ast}$ relation of
galaxies depends on the evolution of the relation between stellar and
molecular masses. \citet{dutton_etal10} find that, in their model, the
effective surface density of atomic hydrogen is $\Shi\approx 10\rm\
M_{\odot}\,yr^{-1}$ and does not evolve with redshift. Our results,
however, indicate that $\Shi$ should increase with increasing
redshift, as metallicities (and, hence, the dust abundance) of
galaxies decrease and their FUV fluxes increase. Conversely,
the $M_{\ast}-M_{\H2}$and SFR-$M_{\ast}$ relations should evolve
differently if their expected dependence on the dust-to-gas ratio and
the FUV flux is taken into account. Given that at lower
metallicities (and, hence, the dust abundance) we expect smaller star
formation rate for the same amount and spatial distribution of neutral
gas, the trends described in this paper may potentially explain why
the model of \citet{dutton_etal10} overpredicts the specific star
formation rate ($\dim{SSFR}\equiv \dim{SFR}/M_{\ast}$) of small-mass
galaxies at $z\gtrsim 3$.

One of the most interesting
results of our simulations is that significant amounts of ionized gas
can be present around high redshift gaseous disks. This ionized gas is
akin to the diffuse ionized gas observed in local galaxies
\citep[e.g.,][]{hoopes_etal03,haffner_etal09} and the Milky Way
\citep{reynolds89,reynolds91,gaensler_etal08}.  Our results indicate
that the ionized gas may dominate the gas mass at low surface
densities ($\Sigma\lesssim 10\Msun\dim{yr}^{-1}$). Furthermore, our
simulations show that ionized gas can remain a significant mass
component at higher gas surface densities in environments with low
dust content and/or high FUV fluxes (e.g., compare gas surface
densities for a given $\Ssfr$ in Figures~\ref{fig:sfltot} and
\ref{fig:sflntr}). One has to keep in mind the possible presence of
significant amounts of ionized gas in theoretical interpretations of
the KS relation and observational estimates of the total gas mass.
The significantly different KS relation in the low dust-to-gas ratio,
high FUV flux environments of high-redshift galaxies may
also strongly bias gas mass estimates that use $z=0$ calibration of
that relation \citep[e.g.,][]{erb_etal06,manucci_etal09}.

As we discussed in \citet{ng:gk10a}, the dust-to-gas ratio
and the FUV flux dependence of the KS relation that we observe
in our simulations has a number of important implications for galaxy
evolution, such as a lower efficiency of star formation in DLA
systems, star formation confined to the highest gas surface densities
of high-$z$ disks, and generally longer gas consumption time scales
in gaseous disks of high-redshift galaxies. The latter can be, at
least partly, responsible for the prevalence of disk-dominated
galaxies at low redshifts. This is because low efficiency of star
formation can maintain disks gas rich until major mergers become rare.
The outer, mostly \emph{gaseous} regions of high-redshift disks
should be more resistant against dynamical heating in mergers
\citep[e.g.,][]{robertson_etal04,robertson_etal06,springel_hernquist05}
and would help maintain forming stellar disks dynamically cold during
minor mergers \citep{moster_etal09b} at later epochs. Moreover, minor
mergers of forming disks should be largely gaseous, and gas brought
in by such mergers should be deposited at large radii as it is ram
pressure stripped by interaction with the gaseous disk and/or halo
around it. This should prevent formation of large bulges, which was
plaguing galaxy formation models, and instead lead to formation of
more extended, higher-angular momentum disks. This scenario is borne
out in recent galaxy formation simulations of \citet*{agertz_etal10},
who show that low efficiency of star formation at high redshifts leads
to more realistic disks and smaller bulge-to-disk ratios.

Another interesting consequence of the complex dependence of the KS
relation on the dust-to-gas ratio and the FUV flux
may be relevant to our own backyard. Recently, \citep{dsh:ogws08}
noted that star formation histories of Milky Way satellites can only
be explained by a KS relation (Equation (\ref{eq:sfitk})) with the sharp
threshold if the threshold varies semi-randomly within a modest
dispersion of about 0.1 dex. This variation is consistent with the
variation given by Equation (\ref{eq:sstar}) for the values of $\D$
and $\U$ typical for dwarf galaxies ($\D\gtrsim0.1$,
$\U\gtrsim1$). Since star formation histories of galactic satellites
are known to be highly variable \citep{dsh:m98,dsh:dwsh05}, the
FUV flux is expected to vary accordingly; such
variations may be responsible for the needed variation of the
threshold in the KS relation, or, more precisely, the characteristic
surface density $\Sigma_{\ast}$ from Equation (\ref{eq:sstar}).

The high mass-to-light ratios (and hence low star formation
efficiencies) of the Local Group dwarf spheroidal galaxies may also be
partially explained by the environmental dependence of $\H2$ abundance
and, hence, star formation. Star formation in such low metallicity,
low dust content dwarf galaxies should be confined only to the highest
gas surface densities (i.e., the central regions) while leaving the
bulk of the gas at lower gas surface densities inert to star
formation. This is consistent with observations of local dwarf low
surface brightness galaxies which exhibit very low molecular gas
fractions and anemic star formation rates
\citep{matthews_etal05,das_etal06,boissier_etal08,wyder_etal09,roychowdhury_etal09}.

The examples described above illustrate the importance of further
investigation of the effects of environmental dependencies of the KS
relation discussed in this paper. The results and fitting formulae
that we present should aid in implementing such dependencies in both
cosmological simulations and semi-analytic models and should thus help
to explore a wide range of possible effects.

\acknowledgements 

This work was supported in part by the DOE at Fermilab, by the NSF
grants AST-0507596 and AST-0708154, and by the Kavli Institute for
Cosmological Physics at the University of Chicago through the NSF
grant PHY-0551142 and an endowment from the Kavli Foundation. The
simulations used in this work have been performed on the Joint
Fermilab - KICP Supercomputing Cluster, supported by grants from
Fermilab, Kavli Institute for Cosmological Physics, and the University
of Chicago. This work made extensive use of the NASA Astrophysics Data
System and {\tt arXiv.org} preprint server.

%----------------------------
\appendix
\section{$\H2$ Formation Model}
%----------------------------

In this Appendix we present the chemical reaction network of hydrogen
and helium, as well as our phenomenological model for the formation of
molecular hydrogen, in full detail \citep[see also][although we note
that the model described here contains some modifications compared to
the model used in this previous paper]{ng:gtk09}.

We follow in detail 8 species of hydrogen and helium: $\HI$,
$\HII$, $\GI$, $\GII$, $\GIII$, $\H2$, $\Hm$, and $\Hp$. It is not,
however, necessary to follow electrons separately, since, in all
physical regimes of interest, abundances of $\Hp$ and $\Hm$ are
extremely small, so
\[
  n_e \approx n_\HII + n_\GII + 2n_{\GIII}.
\]
Note that this equation does not include any negative terms and thus
$n_e$ will always be calculated with the {\it relative\/} error
similar to the relative errors of $n_\HII$, $n_\GII$, and $n_{\GIII}$,
but not larger.

We follow all other species self-consistently and separately by
solving the corresponding ODEs to avoid potentially unbounded increase
of relative error in subtracting abundance of one specie from another
(sometimes called ``loss of precision'').  For example, if the
abundance of $\GIII$ would be calculated by subtracting the abundance
of $\GI$ and $\GII$ from the constant total abundance of He, the
relative error of $\GIII$ can be arbitratily large when the fraction
of $\GIII$ is small.

We explicitly assume that all species are advected with the
same peculiar gas velocity $\vec{v}$. In this case the equations for
the evolution of their number densities can be
concisely represented as 
\begin{equation}
  \frac{\partial n_j}{\partial t} + 3Hn_j + \frac{1}{a}{\rm div}_x(n_j
  \vec{v}) = \dot{\cal I}_j + \dot{\cal M}_j + \dot{\cal D}_j,
  \label{eq:nevol}
\end{equation}
where $j=\HI$, $\HII$, $\GI$, $\GII$, $\GIII$, $\H2$, $\Hm$, and
$\Hp$, the divergence is taken in comoving space $\vec{x}$ and three
terms on the right hand side include reactions due to ionization
balance, molecular chemistry, and dust chemistry respectively. This
subdivision of the reactions into three sets is primarily for the sake
of convenience and because we use different sources for different
reaction rates. This separation is, of course, artificial - all the
reactions take place together in a fluid element.

The OTVET radiative transfer solver produces the radiation field at
each computational cell that is used to calculate the rates for
reactions between chemical species and radiation (including
photo-ionization). We generically label these rates as $\Gamma^{\rm
  RT}$ with various indicies. Since the self-shielding of molecular
hydrogen and shielding by dust are not included in the OTVET solver,
but are the ingredients of our empirical model, they are encapsulated into
two factors, $S_\H2$ and $S_{\rm D}$, with which we multiply the appropriate
rates. These factors are described below.

%-------------------------------
\subsection{Ionization Balance}
%-------------------------------

\def\r#1#2#3{k_{#1}n_{#2}n_{#3}}
\def\prop{\phantom{\bigcap}}

Ionization balance terms include standard processes of
photo-ionization, collisional ionization, and radiative recombination,
and therefore only involve $j=\HI$, $\HII$, $\GI$, $\GII$, $\GIII$. We
label all terms that include at least one of $\H2$, $\Hm$, and $\Hp$
as ``molecular chemistry'', and describe them all in the following
subsection.
\begin{equation}
  \left\{
  \begin{array}{lll}
    \dot{\cal I}_\HI & = &
    -n_\HI\Gamma_\HI - C_\HI n_e n_\HI + R_\HII n_e n_\HII, \\
    \dot{\cal I}_\HII & = &
    -\dot{\cal I}_\HI= - R_\HII n_e n_\HII + n_\HI\Gamma_\HI + C_\HI n_e n_\HI, \\
    \dot{\cal I}_\GI & = &
    -n_\GI\Gamma_\GI - C_\GI n_e n_\GI + (D_\GII+R_\GII) n_e n_\GII, \\
    \dot{\cal I}_\GII & = &
    -n_\GII\Gamma_\GII - (D_\GII+R_\GII) n_e n_\GII - C_\GII n_e n_\GII
    + n_\GI\Gamma_\GI + C_\GI n_e n_\GI + R_\GIII n_e n_\GIII, \\
    \dot{\cal I}_\GIII & = &
    - R_\GIII n_e n_\GIII + n_\GII\Gamma_\GII + C_\GII n_e n_\GII, \\
    \dot{\cal I}_\H2 & = & \dot{\cal I}_\Hm = \dot{\cal I}_\Hp = 0.
  \end{array}
  \right.
  \label{eq:ifull}
\end{equation}
Here $C_j$ are collisional ionization rates, $R_j$ are radiative
recombination rates, and $D_j$ are dielectronic recombination
rates. For these rates we use highly accurate fitting formulae
from \citet{ng:hg97}. The recombination coefficients are computed
self-consistently as a combination of case A and case B recombination,
depending on the gas opacity.

The photo-ionization rates are derived from those
returned by the radiative transfer solver and include the shielding by
dust as
\begin{equation}
  \left\{
  \begin{array}{llll}
    \Gamma_\HI & = & \SD \Gamma^{\rm RT}_\HI & [ \HI+\gamma \rightarrow \HII ], \\
    \Gamma_\GI & = & \SD \Gamma^{\rm RT}_\GI & [ \GI+\gamma \rightarrow \GII ], \\
    \Gamma_\GII & = & \SD \Gamma^{\rm RT}_\GII & [ \GII+\gamma \rightarrow \GIII ].
  \end{array}
  \right.
\end{equation}
In particular, we use the same factor to account for dust shielding in
all three photo-ionization rates. Obviously, this is not exact, as the
dust cross-section is a function of wavelength. However, since the effect
of helium on molecular chemistry inside molecular clouds is thought to
be small, helium ionization inside molecular clouds is sufficient to be
treated rather approximately.

%-------------------------------
\subsection{Molecular Chemistry}
%-------------------------------

Molecular chemistry terms include a large set of reactions between
$\H2$, $\Hp$, and $\Hm$ and atomic species. The full set of equations we call
{\it ``the full 8-species Model''}:
\begin{equation}
  \left\{
  \begin{array}{lll}
    \dot{\cal M}_\HI & = &
    \GA n_\Hm + \GB n_\Hp + 2 \GE n_\H2 + 2
    \GLW n_\H2 - \r{1}e\HI - \r{2}\Hm\HI -
    \r{3}\HII\HI - \r{4}\Hp\HI - \\
    & & \r{26}\GII\HI - 2k_{30}n_\HI^3 - 2k_{31}n_\HI^2n_\H2 -
    2k_{32}n_\HI^2n_\GI + 2\r{5}\HII\Hm + 2\r{6}e\Hp + \r{7}{\H2}\HII + \\
    & & 2\r{8}e{\H2} + 2\r{9}\HI{\H2} + 2\r{10}{\H2}{\H2} + 2\r{11}\GI{\H2} +
    \r{14}e\Hm + \r{15}\HI\Hm + \r{21}\Hp\Hm + \\
    & & 3\r{22}\Hm\Hp + \r{23}e{\H2} + \r{24}\GII{\H2} + \r{27}\GI\HII
    + \r{28}\GII\Hm + \r{29}\GI\Hm, \\\def\GD{\Gamma_{\rm D}}
    \dot{\cal M}_\HII & = &
    \GB n_\Hp + 2 \GC n_\Hp
    -\r{3}\HI\HII -\r{5}\Hm\HII - \r{7}{\H2}\HII - \r{16}\Hm\HII -
    \r{27}\GI\HII + \r{4}\Hp\HI + \\
    & & \r{24}\GII{\H2} + \r{26}\HI\GII, \\
    \dot{\cal M}_\GI & = &
    -\r{27}\HII\GI - \r{29}\Hm\GI
    + \r{24}\GII{\H2} + \r{25}\GII{\H2} + \r{26}\GII\HI +
    \r{28}\GII\Hm, \\
    \dot{\cal M}_\GII & = &
    -\r{24}{\H2}\GII - \r{25}{\H2}\GII - \r{26}\HI\GII - \r{28}\Hm\GII
    +\r{27}\HII\GI + \r{29}\Hm\GI, \\
    \dot{\cal M}_\GIII & = & 0, \\
    \dot{\cal M}_\H2 & = &
    -\GD n_\H2 - \GE n_\H2 - \GLW n_\H2 -\r{7}{\H2}\HII -
    \r{8}e{\H2} - \r{9}\HI{\H2} - \r{10}{\H2}{\H2} - \r{11}\GI{\H2} -
    \\ 
    & & \r{23}e{\H2} -\r{24}\GII{\H2} - \r{25}\GII{\H2} + \r{2}\Hm\HI
    + \r{4}\Hp\HI + \r{21}\Hp\Hm + k_{30}n_\HI^3 + \\
    & & k_{31}n_\HI^2n_\H2 + k_{32}n_\HI^2n_\GI, \\ 
    \dot{\cal M}_\Hp & = &
    -\GB n_\Hp - \GC n_\Hp + \GD n_\H2 -\r{4}\HI\Hp
    - \r{6}e\Hp - \r{21}\Hm\Hp - \r{22}\Hm\Hp + \r{3}\HI\HII + \\ 
    & & \r{7}{\H2}\HII +\r{16}\HII\Hm + \r{25}{\H2}\GII, \\
    \dot{\cal M}_\Hm & = &
    -\GA n_\Hm - \r{2}\HI\Hm - \r{5}\HII\Hm
    - \r{14}e\Hm - \r{15}\HI\Hm - \r{16}\HII\Hm - \r{21}\Hp\Hm - \\
    & & -\r{22}\Hp\Hm - \r{28}\GII\Hm - \r{29}\GI\Hm + \r{1}e\HI +
    \r{23}e{\H2}, 
  \end{array}
  \right.
  \label{eq:mfull8}
\end{equation}
where 
\begin{equation}
  \left\{
  \begin{array}{llll}
    \GA & = & \SD \Gamma^{\rm RT}_{\rm A} & [ \Hm + \gamma \rightarrow \HI + e ], \\
    \GB & = & \SD \Gamma^{\rm RT}_{\rm B} & [ \Hp + \gamma \rightarrow \HI + \HII ], \\
    \GC & = & \SD \Gamma^{\rm RT}_{\rm C} & [ \Hp + \gamma \rightarrow 2\HII+ e ], \\
    \GD & = & \SD S_\H2 \Gamma^{\rm RT}_{\rm D} & [ \H2 + \gamma \rightarrow \Hp + e ], \\
    \GE & = & \SD S_\H2 \Gamma^{\rm RT}_{\rm E} & [ \H2 + \gamma \rightarrow 2\HI~~(h\nu > 13.6\dim{eV}) ], \\
    \GLW & = & \SD S_\H2 \Gamma^{\rm RT}_{\rm LW} & [ \H2 + \gamma \rightarrow 2\HI~~(\mbox{Lyman-Werner band}) ].
  \end{array}
  \right.
  \label{eq:pmrates}
\end{equation}
The rate coefficients $k_1$-$k_{32}$ are taken from 
\citet{ism:ga08}; we do not list here all these reactions for
brevity. Cross sections for photo-rates A-D are given by
\citet{ism:sk87}, while the cross section for the reaction E is given
by \citet{ism:aazn97}, for both ortho- and para-$\H2$. The radiative
transfer in the Lyman-Werner bands $\Gamma^{\rm RT}_{LW}$ is treated
fully self-consistently with 20{,}000 frequency bins, as described in
\citet{ng:rgs02a}.

Analogously to the previous section, we use the same $S_\H2$ factor to
account for $\H2$ self-shielding for reactions D, E, and LW. This is a
crude approximation, but a more accurate treatment would introduce
additional parameters that cannot yet be calibrated with the existing
limited observational measurements.

Equations (\ref{eq:mfull8}) can be substantially simplified if we note that in all physical regimes relevant to cosmology the
abundances of $\Hp$ and $\Hm$ are always extremely small, so that they
can always be assumed to be in the kinetic equilibrium, $\dot{\cal
  M}_\Hp \approx \dot{\cal M}_\Hm \approx 0$ (T.\ Abel, private 
communication). With this assumption and neglecting reactions involving $k_{21}$ and $k_{22}$, because their rates are
$\propto n_\Hm n_\Hp$ where both $n_\Hm$ and $n_\Hp$ are small, 
expressions for the equilibrium abundances of $\Hp$ and $\Hm$ can be
derived in a closed form, resulting in the following {\it ``6-species
  model''}: 
\begin{equation}
  \left\{
  \begin{array}{lll}
    n_\Hm & = & \frac{\displaystyle\prop
      \r{1}e\HI + \r{23}e{\H2}}
    {\displaystyle\prop
      \GA + k_{2}n_\HI + k_{5}n_\HII + k_{14}n_e + k_{15}n_\HI
      + k_{16}n_\HII + k_{28}n_\GII + k_{29}n_\GI}, \\ 
%
%    n_\Hm & = & (\r{1}e\HI + \r{23}e{\H2})/(\GA + k_{2}n_\HI +
%    k_{5}n_\HII + k_{14}n_e + k_{15}n_\HI + k_{16}n_\HII +
%    k_{28}n_\GII + k_{29}n_\GI), \\ 
%
    n_\Hp & = & \frac{\displaystyle\prop
      \GD n_\H2 + \r{3}\HI\HII + \r{7}{\H2}\HII +\r{16}\HII\Hm +
      \r{25}{\H2}\GII} 
    {\displaystyle\prop
      \GB + \GC + k_{4}n_\HI + k_{6}n_e}, \\
    \dot{\cal M}_\HI & = &
    \GA n_\Hm + \GB n_\Hp + 2 \GE n_\H2 + 2
    \GLW n_\H2 - \r{1}e\HI - \r{2}\Hm\HI -
    \r{3}\HII\HI - \r{4}\Hp\HI - \\
    & & \r{26}\GII\HI - 2k_{30}n_\HI^3 - 2k_{31}n_\HI^2n_\H2 -
    2k_{32}n_\HI^2n_\GI + 2\r{5}\HII\Hm + 2\r{6}e\Hp + \r{7}{\H2}\HII + \\
    & & 2\r{8}e{\H2} + 2\r{9}\HI{\H2} + 2\r{10}{\H2}{\H2} + 2\r{11}\GI{\H2} +
    \r{14}e\Hm + \r{15}\HI\Hm + \\
    & & \r{23}e{\H2} + \r{24}\GII{\H2} + \r{27}\GI\HII
    + \r{28}\GII\Hm + \r{29}\GI\Hm,  \\
    \dot{\cal M}_\HII & = &
    \GB n_\Hp + 2 \GC n_\Hp
    -\r{3}\HI\HII -\r{5}\Hm\HII - \r{7}{\H2}\HII - \r{16}\Hm\HII -
    \r{27}\GI\HII + \r{4}\Hp\HI + \\
    & & \r{24}\GII{\H2} + \r{26}\HI\GII, \\
    \dot{\cal M}_\GI & = &
    -\r{27}\HII\GI - \r{29}\Hm\GI + \r{24}\GII{\H2} + \r{25}\GII{\H2}
    + \r{26}\GII\HI + \r{28}\GII\Hm, \\
    \dot{\cal M}_\GII & = &
    -\r{24}{\H2}\GII - \r{25}{\H2}\GII - \r{26}\HI\GII - \r{28}\Hm\GII
    +\r{27}\HII\GI + \r{29}\Hm\GI, \\
    \dot{\cal M}_\GIII & = & 0, \\
    \dot{\cal M}_\H2 & = &
    -\GD n_\H2 - \GE n_\H2 - \GLW n_\H2 -\r{7}{\H2}\HII -
    \r{8}e{\H2} - \r{9}\HI{\H2} - \r{10}{\H2}{\H2} - \r{11}\GI{\H2} - \\
    & & \r{23}e{\H2} -\r{24}\GII{\H2} - \r{25}\GII{\H2} + \r{2}\Hm\HI
    + \r{4}\Hp\HI + k_{30}n_\HI^3 + \\
    & & k_{31}n_\HI^2n_\H2 + k_{32}n_\HI^2n_\GI. 
  \end{array}
  \right.
  \label{eq:mfull6}
\end{equation}

Finally, under normal ISM conditions the ionization balance of
hydrogen and helium is controlled by the radiative recombination, 
photo-ionization and ionization by cosmic rays. In this limit
we can ignore all gas-phase molecular chemistry reactions,
\[
  \dot{\cal M}_j \approx 0.
\]
We dub this approximation the {\it ``minimal model''}. The minimal
model is often (justifiably) used in studies of local ISM
\citep[c.f.][]{sfr:km05,pelupessy_etal06,sfr:kt07,ism:gm07a,ism:gm07b},
but is also occasionally applied to high-redshift or low-metallicity
systems \citep{sfr:kept09,pelupessy_popadopoulos09}. We find, however,
that the minimal model produces results that are reasonably close to
the full model for $\D \gtrsim 0.1$ (for any FUV flux), but
becomes progressively less accurate for lower dust-to-gas ratios,
mis-predicting the atomic-to-molecular transition as a function of
density by a factor of 2 for $\D \sim 0.01$.

In order to maintain high accuracy for the full sampled range of $\D$
and $\U$, \emph{all simulations presented in this paper were performed
with the 6-species model}.

\subsection{Dust Chemistry}

In our model the only dust chemistry reaction that we include is the
formation of molecular hydrogen on dust,
\begin{equation}
  \left\{
  \begin{array}{lll}
    \dot{\cal D}_\H2 & = & \D R_0 C_\rho n_\HI (n_\HI + 2 n_\MH), \\
    \dot{\cal D}_\HI & = & -2\dot{\cal D}_\H2, \\
    \dot{\cal D}_\HII & = & \dot{\cal D}_\GI = \dot{\cal D}_\GII = \dot{\cal D}_\GIII = \dot{\cal D}_\Hm = \dot{\cal D}_\Hp = 0,
  \end{array}
  \right.
  \label{eq:dfull}
\end{equation}
where $R_0=3.5\times10^{-17}\dim{cm}^3\,\dim{s}^{-1}$ \citep[][see Equation (\ref{eq:pardefs})]{ism:wthk08} and
$C_\rho$ is the clumping factor inside molecular clouds, which takes
into account the fact that the gas is clumped on subgrid scales
unresolved in our simulations \citep[also see][]{ng:gtk09}. The
clumping factor $C_\rho$ is a parameter of our model, we discuss a
reasonable choice for its value below, in \S \ref{sec:calib}.

\subsection{Heating, cooling, and thermodynamics}
\label{sec:hetacool}

For the heating and cooling terms in the equation for the internal
energy we include all of the terms normally included in the simulations
of first stars and in the ISM models. Specifically, the entropy term
in the energy equation for the gas can be written as
\[
  \rho T \frac{ds}{dt} = \dot{\cal H} - \dot{\cal C},
\]
where $\dot{\cal H}$ and $\dot{\cal C}$ are heating and cooling terms,
\begin{align}
  \dot{\cal H} & = \dot{\cal H}_{\rm PI} + \dot{\cal H}_{\rm CMB} + \dot{\cal H}_{\rm
  Ly\alpha} + \dot{\cal H}_{\H2} + \dot{\cal H}_{\rm PAH} + \dot{\cal H}_{\rm CR},
  \nonumber \\
  \dot{\cal C} & = \dot{\cal C}_{\rm CI} + \dot{\cal C}_{\rm RR} + \dot{\cal C}_{\rm
  DER} + \dot{\cal C}_{\rm LE, A} + \dot{\cal C}_{\rm
  FF} + \dot{\cal C}_{\rm QX} + \dot{\cal C}_{\rm LE, \H2} + \dot{\cal C}_{{\rm LE,}
  Z} + \dot{\cal C}_{\rm D}.
\end{align}

In the heating function, we include
\begin{description}
\item[$\dot{\cal H}_{\rm PI}$]: photoionization heating due to $\HI$,
  $\GI$, and $\GII$, using cross-sections from \citet{ng:hg97};
\item[$\dot{\cal H}_{\rm CMB}$]: Compton heating/cooling on the CMB \citep{ng:hg97};
\item[$\dot{\cal H}_{\rm Ly\alpha}$]: heating by Ly$\alpha$ photons \citep{igm:tmmr00};
\item[$\dot{\cal H}_{\rm \H2}$]: heating due to photo-dissociation of
  $\H2$, $\dot{\cal H}_{\rm \H2} =
  0.4\dim{eV}\times n_\H2\left(\GD+\GE+\GLW\right)$
  (Equation (\ref{eq:pmrates}));
\item[$\dot{\cal H}_{\rm PAH}$]: photo-electric heating on PAH, implemented
  as in \citet{ism:gm07a};
\item[$\dot{\cal H}_{\rm CR}$]: cosmic rate heating, assuming that the
  cosmic rate density scales as the dust-to-gas ratio, implemented
  as in \citet{ism:gm07a}.
\end{description}

Cooling processes include 
\begin{description}
\item[$\dot{\cal C}_{\rm CI}$]: cooling due to collisional
 ionizations of $\HI$, $\GI$, and $\GII$ \citep{ng:hg97};
\item[$\dot{\cal C}_{\rm RR}$]: cooling due to radiative
  recombinations of $\HII$, $\GII$, and $\GIII$ \citep{ng:hg97};
\item[$\dot{\cal C}_{\rm DER}$]: cooling due to di-electronic
  recombination of $\GIII$ \citep{ng:hg97};
\item[$\dot{\cal C}_{\rm LE, A}$]: line exitation cooling of $\HI$ and
  $\GII$ \citep{ng:hg97};
\item[$\dot{\cal C}_{\rm FF}$]: free-free emission \citep{ng:hg97};
\item[$\dot{\cal C}_{\rm QX}$]: cooling due to charge exchange
  reactions between $\H2$, $\Hm$, $\HI$ and free electrons (reactions
  8, 9, 10, 14, and 15 from \citet{ism:ga08});
\item[$\dot{\cal C}_{\rm LE, \H2}$]: line exitation cooling of $\H2$
  \citep{ism:ga08};
\item[$\dot{\cal C}_{{\rm LE,} Z}$]: line exitation cooling of heavy
  elements, using \citet{atom:sd93} cooling functions for
  $T>10^4\dim{K}$ and \citet{atom:p70} and \citet{atom:dm72} rates in
  the $T<10^4\dim{K}$ regime;
\item[$\dot{\cal C}_{\rm D}$]: cooling on dust from \citet{atom:d81}.
\end{description}

Some of the reaction rates involving $\H2$ depend on the ortho-to-para
ratio of molecular hydrogen. For this ratio and other thermodynamic
quantities ($\gamma(T)$, $U(T)$, etc) we use exact expressions
computed from quantum-mechanical statistical sums (Turk et al, 2010,
in preparation). 

\begin{figure}[t]
\includegraphics[scale=0.90]{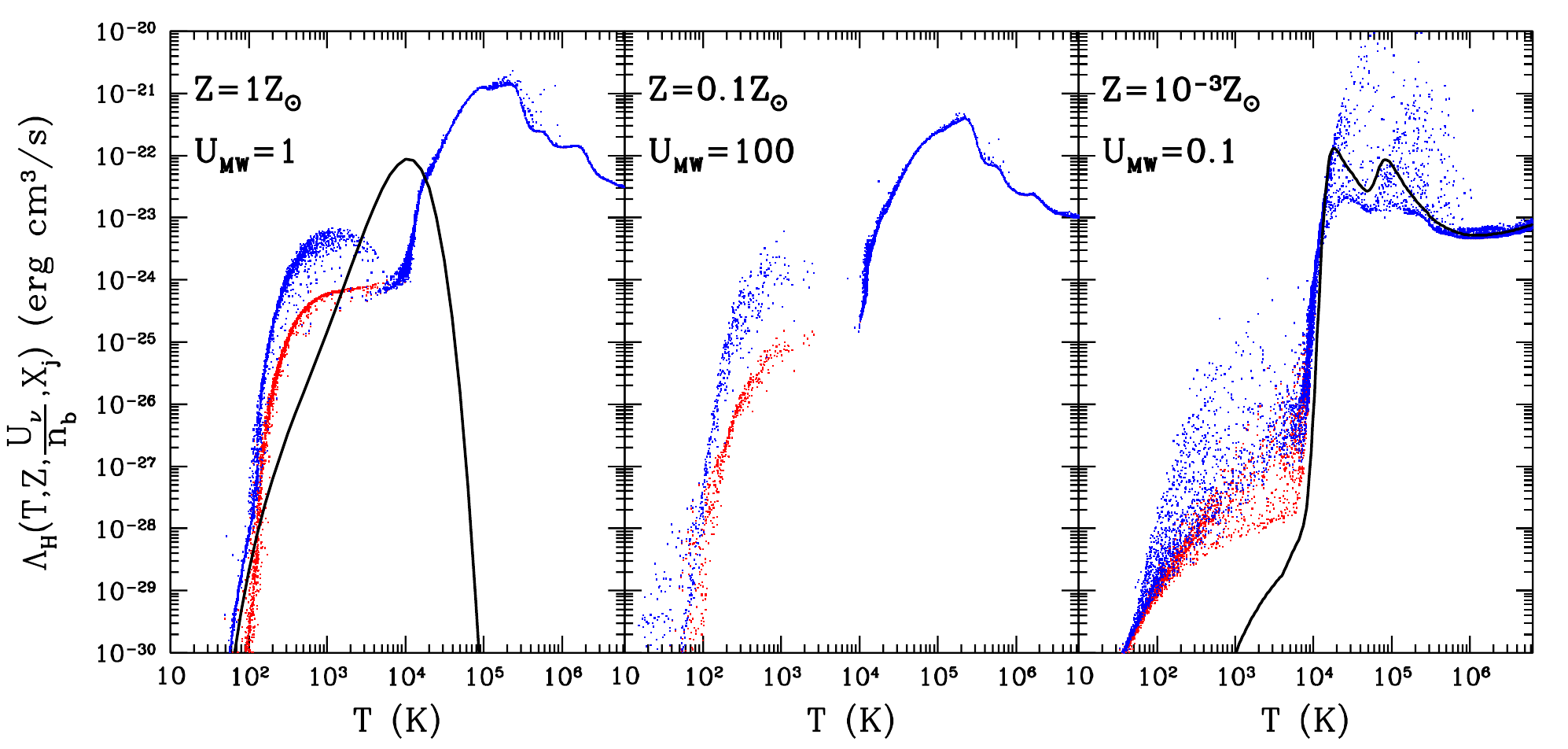}
\vspace{0.5cm}
\caption{Cooling functions (per hydrogen nuclesus) for 3
  representative values of gas metallicity $Z$ and the FUV 
  flux $\U$. In this plot we assume $\D=Z/Z_\odot$. Blue points show
  the full cooling function (including all relevant physical
  processes), while red points show the result of excluding $\H2$
  cooling. Black lines trace the $\H2$ cooling function from
  \citet{ism:gp98} (left panel) and the standard, metal-free cooling
  function (right panel).\label{fig:cf}\newline}
\end{figure}

Examples of cooling functions from our simulations are given in Figure
\ref{fig:cf}. The cooling function, in general, is \emph{not} a
function of gas temperature only, but also depends on the gas
metallicity $Z$, the energy density of the incident radiation field
$U_\nu$, the number density of baryons $n_b$ (although for
$n_b\lesssim10^4\dim{cm}^{-3}$ the dependence on the last two
parameters always enters as $U_\nu/n_b$), and abundances of all atomic
and molecular species $X_j\equiv n_j/n_b$. Therefore, when plotted as
a function of temperature, the cooling function takes a range of
values (depending on the values of other gas properties) rather than a
single, unique value.

Interestingly, Figure \ref{fig:cf} shows that the cooling rate at
$T<10^4\dim{K}$ is dominated by cooling due to molecular hydrogen,
rather than by low ionization metal species such as OI or
CII. Molecular hydrogen cooling is often assumed to be negligible
\citep[c.f.][]{ism:wmh03,stahler_palla05} due to lower cooling rates
\citep[c.f.][]{ism:gp98}. However, we use the updated $\H2$ cooling rates  
of \citet{ism:ga08}, which are considerably higher than the previous
estimates.  As Figure~\ref{fig:cf} shows, the new $\H2$ cooling rates
dominate over the low ionization metal species at $T\lesssim 5000\dim{K}$.

%------------------------------
\subsection{Shielding Factors}
%------------------------------

The two shielding factors, $\SD$ and $S_\H2$, together with the
clumping factor $C_\rho$, are important parameters of our empirical
model. As \citet{ng:gtk09} explain, we use an ansatz similar in spirit
to the Sobolev approximation to estimate dust shielding:
\begin{equation}
  \SD = e^{\displaystyle - \D \sigma_0 (n_\HI + 2 n_\MH) L_{\rm Sob}},
  \label{eq:sd}
\end{equation}
where $\D$ is the dust-to-gas ratio in units of its Milky Way value
(see \S~\ref{sec:sims}), $\sigma_0 = 2\times 10^{-21} \dim{cm}^2$, and 
\begin{equation}
  L_{\rm Sob} \equiv \rho/(2 |\nabla\rho|).
  \label{eq:sob}
\end{equation}
Note that the value for $\sigma_0$ that we
use in this paper is twice lower than the one listed in
\citet{ng:gtk09}; the new value is a commonly adopted value for this
parameter for the Milky Way type dust, and provides a better
quantitative fit to the existing 
observational constraints. In addition, a factor of 2 in the
denominator of the expression for $L_{\rm Sob}$ was missing in
\citet{ng:gtk09} - this was a typo, and the correct expression was
used when simulations were run. 

The major change between our current model and the model of
\citet{ng:gtk09} is in the form of the molecular hydrogen
self-shielding factor. In \citet{ng:gtk09} this form  was modified from the commonly used formula of
\citet{ism:db96}, because the FUV flux in \citet{ng:gtk09} was
much higher than the Draine value. In our present tests, we find that we
can use either the original \citet{ism:db96} formula or their simpler and more
approximate expression,
\begin{equation}
  S_\H2 = \left\{
  \begin{array}{ll}
    1, & \mbox{for } N_\H2 < 10^{14}\dim{cm}^{-2}, \\
    \left(N_\H2/10^{14}\dim{cm}^{-2}\right)^{-3/4}, & \mbox{for }
    N_\H2 > 10^{14}\dim{cm}^{-2}, 
  \end{array}
  \right.
  \label{eq:sh2}
\end{equation}
which we actually use for computational efficiency\footnote{We have
  indeed verified that a more complex formula (Equation (37) of
  \citet{ism:db96}) produces essentially indistinguishable results from
  the more approximate form of Equation (\ref{eq:sh2}).}. 

Finally, to complete the full specification of our chemical model, we
need to estimate the column density of the molecular gas, $N_\H2$, for
the self-shielding factor given by Equation (\ref{eq:sh2}). Unfortunately,
we cannot simply use the Sobolev approximation to derive $N_\H2$
similar to the column density of dust in Equation (\ref{eq:sd}),
because $\H2$ absorption is concentrated in separate absorption lines
and is sensitive to the internal velocity dispersion inside molecular
clouds. These velocities are unresolved in our simulations, but can
greatly reduce the self-shielding of molecular gas.  Dust, on the
other hand, absorbs UV radiation in continuum and is thus not affected
by velocity distribution of the gas.

Therefore, we introduce the following simple ansatz for the effective
column density $N_\H2$ for Equation (\ref{eq:sh2}),
\begin{equation}
  N_\H2 \approx n_\H2 L_c,
  \label{eq:lcdef}
\end{equation}
where $L_c$ is the velocity coherence length of the molecular hydrogen
inside molecular clouds. Since we cannot deduce this quantity from
observations or other calculations, we treat it as another parameter
of our model.

With the expressions for the shielding factors above, the only two
parameters of our model are $C_\rho$ and $L_c$. These parameters can
only be determined by comparing the simulation results to the
observational data.

\subsection{Calibration}
\label{sec:calib}

As the primary data sets used to calibrate the model, we use the
measurements of atomic and molecular gas surface densities in nearby
spirals from \citet{misc:wb02} and measurements of gas fractions along
the lines of sight to individual stars for atomic \citep{ism:gl05}
and molecular gas in the Milky Way and Magellanic Clouds
\citep{ism:tsrb02,ism:gstd06,ism:wthk08}. 

\begin{figure}[t]
\includegraphics[scale=0.45]{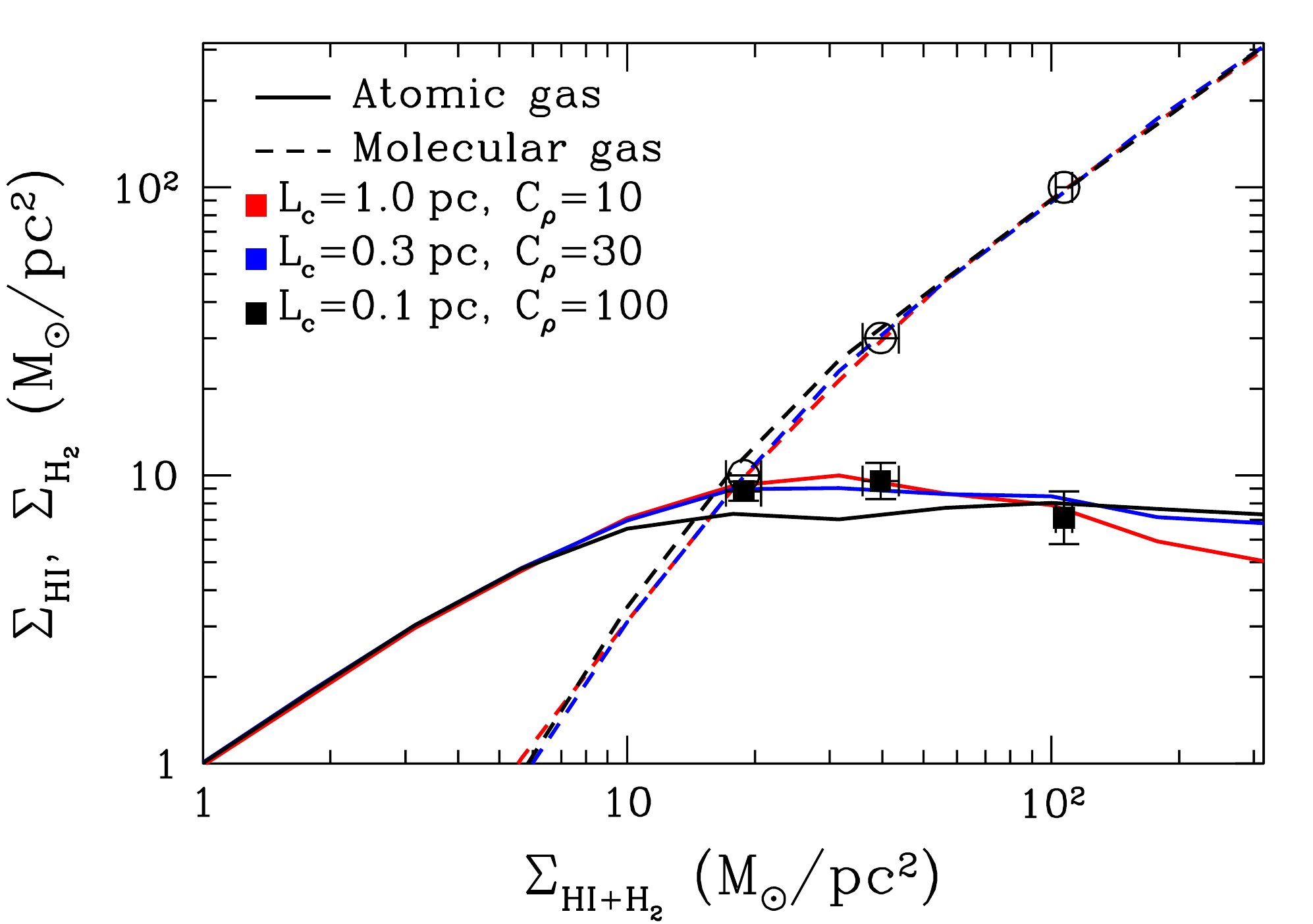}
\includegraphics[scale=0.45]{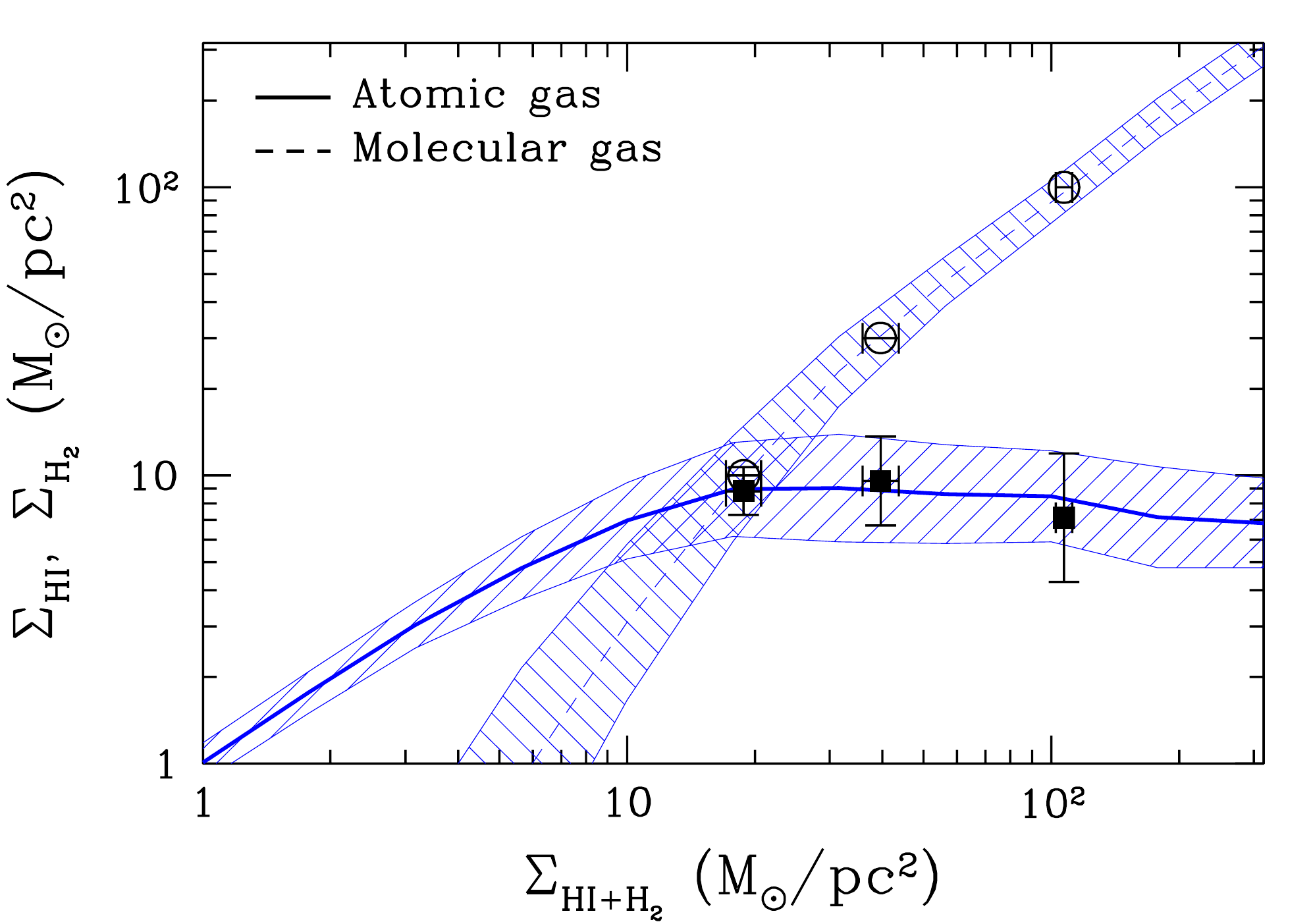}
%\epsscale{1.15}
%\plottwo{\figname{shish2a.ps}}{\figname{shish2b.ps}}
\caption{Average atomic and molecular gas surface densities as
  functions of the total (neutral) hydrogen gas surface density 
  averaged over $500\dim{pc}$ scale. The left panel show three test
  simulations with three values of the clumping factor $C_\rho$ and
  molecular coherence length $L_c$.  Filled squares and open circles
  with error bars mark the observed average atomic and molecular
  hydrogen surface densities at $\Smol=10$, $30$, and
  $100\Msun\dim{pc}^{-2}$ from \citet{misc:wb02}. The right panel shows
  our fiducial model ($L_c=0.3\dim{pc}$,$C_\rho=30$) together with the
  rms scatter (shaded bands) around the averages. The error-bars on
  the observational points now show the dispersion around the average
  rather than the error of the mean.\label{fig:asjust}\newline}
\end{figure}

We calibrate the two parameters of the model: the clumping factor $C_\rho$
and the molecular coherence length $L_c$. We find, however, that there
is no unique best-fit set of parameters. Instead, any combination of
these two parameters that satisfy the constraint 
\[
  L_cC_\rho \approx 10\dim{pc}
\]
provides an acceptable fit to the observational constraints. As an
example, we show on the left panel of Figure \ref{fig:asjust} fits to
the \citet{misc:wb02} measurements (averaged over all galaxies they
observed) for three combinations of the parameters $L_c$ and $C_\rho$. In
general, higher clumping factors result in the lower atomic contents
at high surface densities, but the trend is too weak to be of any
statistically significant constraining power.

\begin{figure}[t]
\includegraphics[scale=0.45]{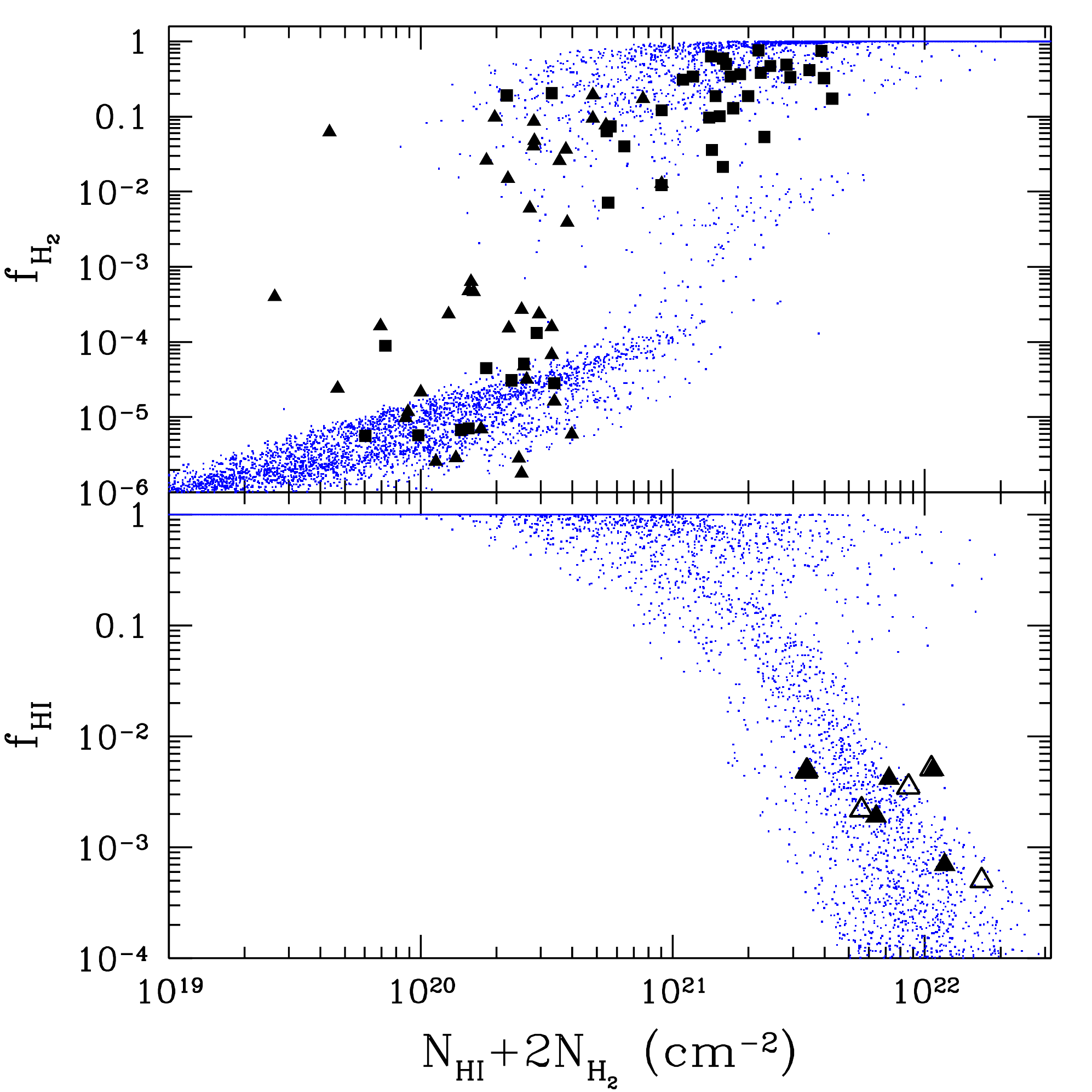}%
\includegraphics[scale=0.45]{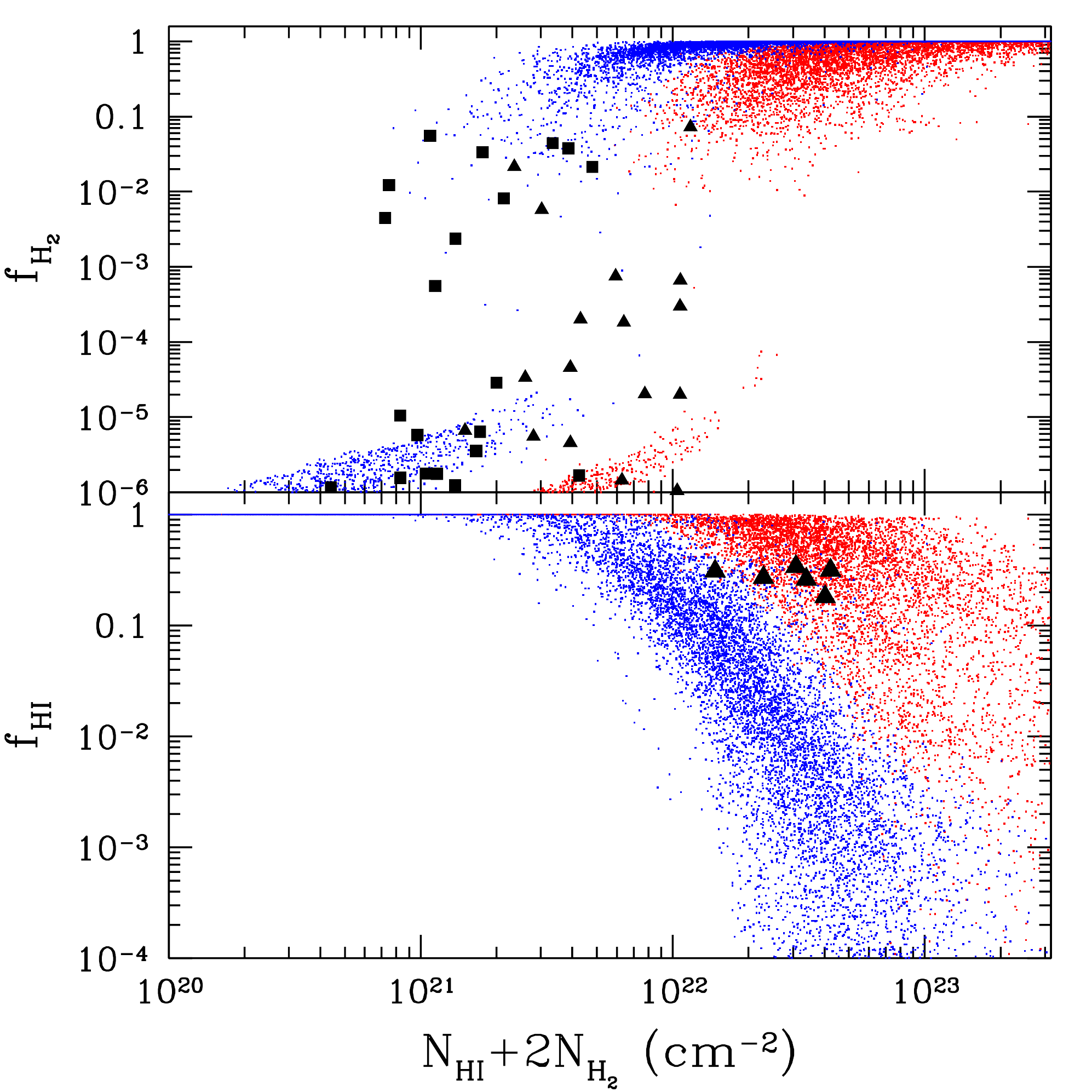}
%\epsscale{1.15}
%\plottwo{\figname{fhih2a.ps}}{\figname{fhih2b.ps}}
\caption{Atomic (bottom) and molecular (top) gas fractions as
  functions of the total (neutral) hydrogen gas column density along
  individual lines of sight through the galactic disks. Colored points
  shows our fiducial test simulation ($L_c=0.3\dim{pc}$, $C_\rho=30$),
  while black points show observational measurements. The left panel
  shows the $(\D=1,\U=1)$ simulation case and the observational
  measurements of molecular fractions in the Milky Way galaxy from 
  \citet{ism:gstd06} (filled triangles) and \citet{ism:wthk08} (filled
  squares) and atomic fractions measurements from \citet{ism:gl05}. The
  right panel shows $(\D=0.3,\U=10)$ (blue points) and
  $(\D=0.1,\U=100)$ (red points) simulation cases that should bracket
  possible values of these parameters for Magellanic Clouds. Filled
  saquares and triangles on the top panel show the measurements for
  LMC and SMC molecular fractions respectively \citep{ism:tsrb02}. On
  the bottom panel the measurements are for SMC \citep{ism:lbsm07}, to
  be compared with red points.\label{fig:calib}}
\end{figure}

As a fiducial set of parameters we choose the combination
$L_c=0.3\dim{pc}$ and $C_\rho=30$. This choice provides a marginally
better overall fit to the observations, and is also consistent with
estimates of the gas clumping factor deep inside molecular clouds
\citep{sfr:mo07}. The fiducial value of $C_\rho$ is somewhat larger
than the estimates of the clumping factor from numerical simulations
of turbulent molecular clouds, $C_\rho = e^{\sigma_{\ln\rho}^2}$,
where $\sigma_{\ln\rho}\approx 1 - 1.5$ is the dispersion of the
lognormal density distribution inside the clouds. However, the value
of $C_\rho=10$, which was used in \citet{ng:gtk09} and is more
consistent with the numerical simulations of turbulent molecular
clouds would provide an almost equally good to the existing
observations, if it is used with $L_c\approx1\dim{pc}$.

\subsection{Dependence on Numerical Resolution}

\begin{figure}[t]
\includegraphics[scale=0.45]{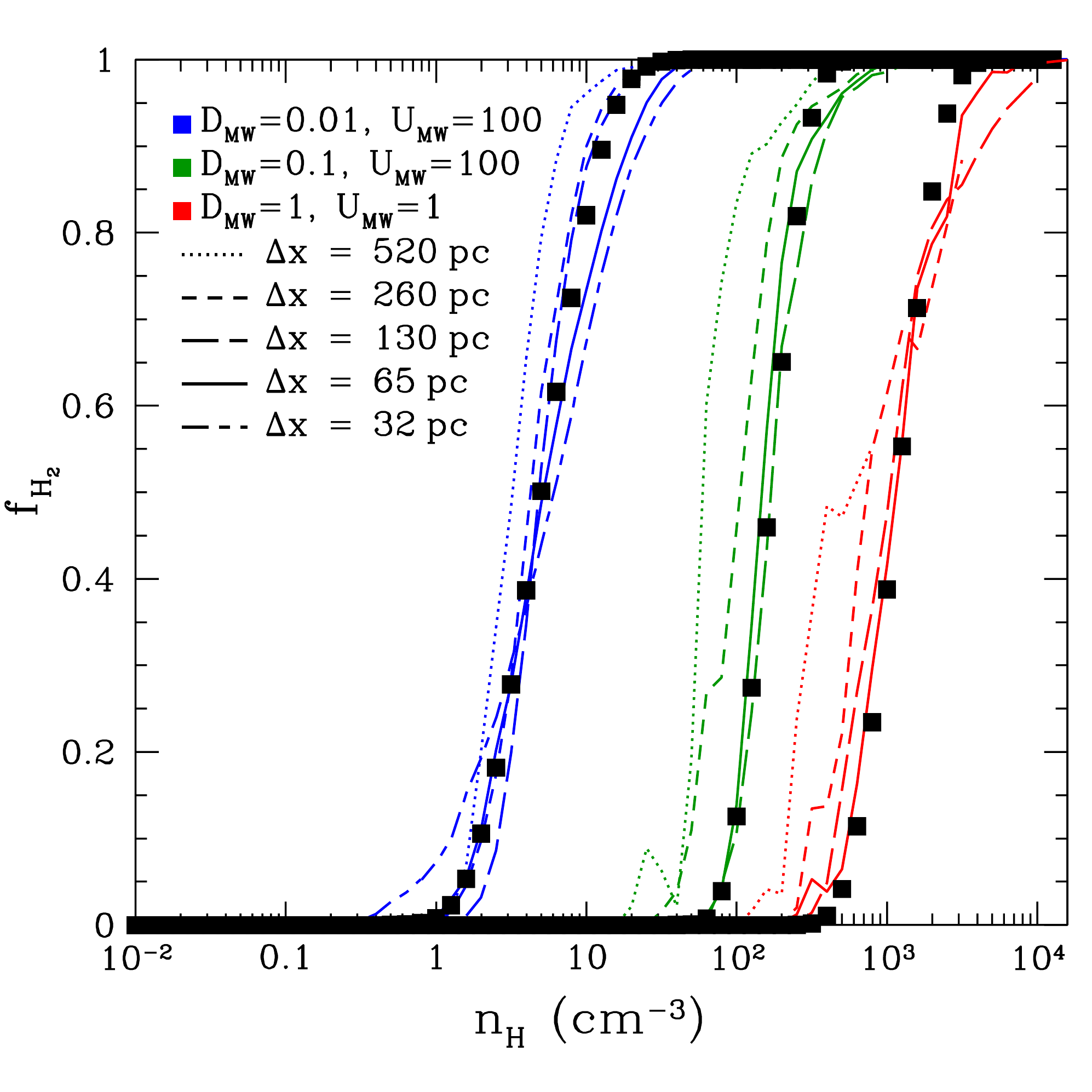}%
\includegraphics[scale=0.45]{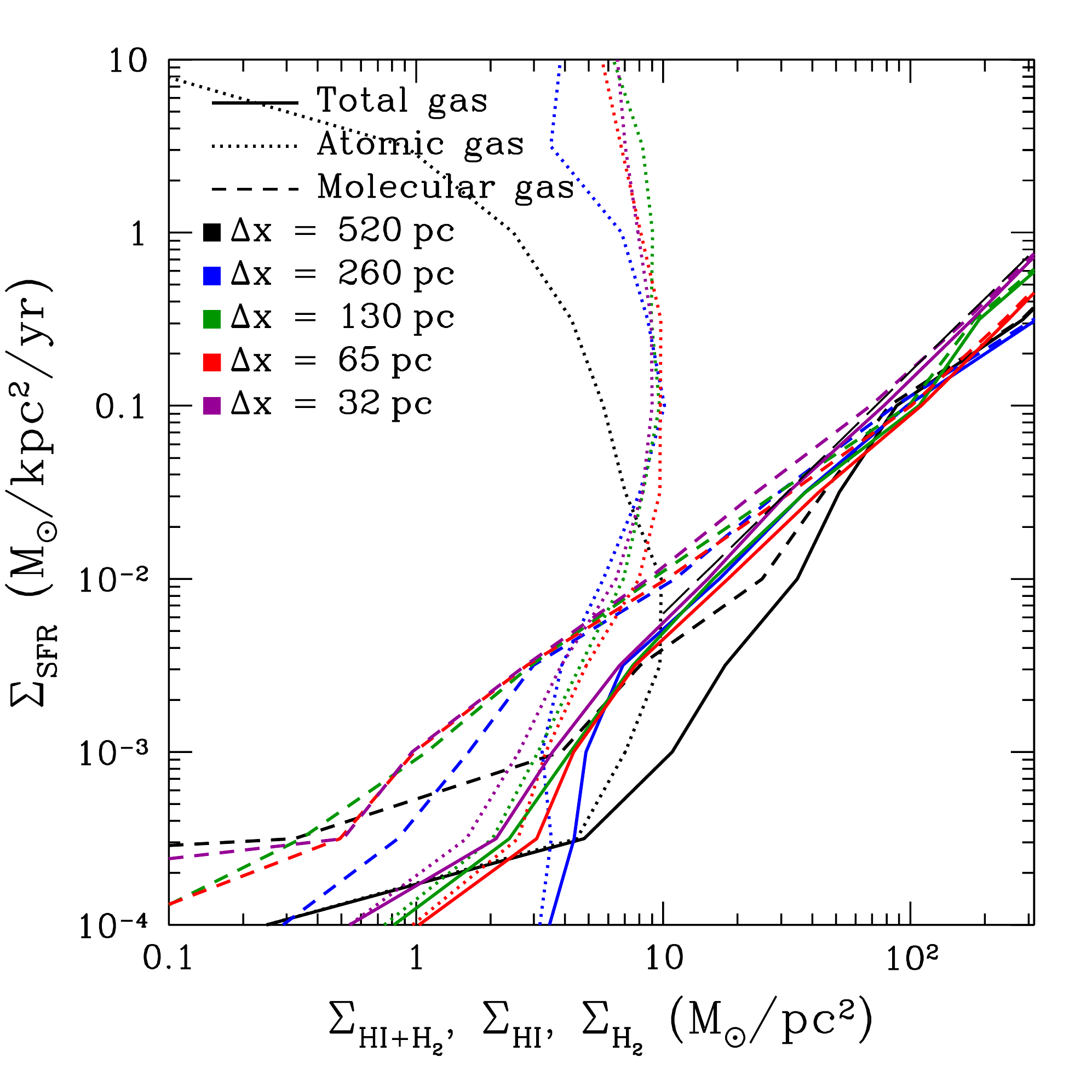}
%\epsscale{1.15}
%\plottwo{\figname{convfh2.ps}}{\figname{convsfl.ps}}
\caption{\label{fig:conv}
Dependence of the atomic-to-molecular transition (left) and the KS
relation (right) on numerical resolution in our model. The left panel
shows three representativel cases $(\D,\U)=(1,1)$,
$(\D,\U)=(0.1,100)$, and $(\D,\U)=(0.01,100)$, while only the first
case (Milky Way like parameters) is shown on the right panel for the
sake of clarity (the other two cases show similar behavior). The value
of the cell size $\Delta x$ on the highest resolved level is shown for
each line. Black squares on the left panel trace the approximate fit
(\ref{eq:fh2fit}).}
\end{figure}

Any sub-cell model would be of limited value, if it was only
applicable to a narrow range of numerical resolutions. In order to
test the range of spatial resolutions over which our model performs
robustly, we have re-run a subset of our test simulations, varying the
maximum allowed level of refinement between 6 and 10, compared to our
fiducial value of 9 (cell size of $\Delta x=65\dim{pc}$ at $z=3$ in
physical units).

The results of these tests are shown in Figure \ref{fig:conv} for the
atomic-to-molecular transition and the KS relation. In order to
perform a genuine resolution test, in each run with different
resolution we only show cells that are refined to the lowest allowed
level. For example, in the run with the maximum level 10, we only show
cells from level 10, so that level 9 cells, which are also present in
that test run, do not contaminate Fig.\ \ref{fig:conv}. Of course, in
realistic simulations cells from all levels that contain molecular gas
are going to contribute to the $f_\H2 - n_\Ht$ relation, so Fig.\
\ref{fig:conv} actually \emph{exaggerates} the effect of changing
resolution.
 At resolutions
$\Delta x \la 260\dim{pc}$ our model performs robustly down to the
smallest scales we are able to probe ($\Delta x \approx
30\dim{pc}$). At coarser resolution of $\Delta x=520\dim{pc}$ small
molecular clouds in low density gas are not captured properly,
resulting in a sharper fall-off in the KS relation at low values of
$\Sntr$. In addition, the Sobolev-like approximation for the dust
column density (Equation (\ref{eq:sob})) overestimates the column density
significantly, which results in the atomic-to-molecular transition
shifting towards lower density gas (especially for low dust-to-gas
ratio and high FUV flux). We conclude, therefore, that
spatial resolution of at least $250\dim{pc}$ is required for our model
to work robustly.  

%\begin{figure}[t]
%\includegraphics[scale=0.45]{convfit1}%
%\includegraphics[scale=0.45]{convfit2}
%\caption{\label{fig:convfit}
%The dependence of our fitting formula (Equation (\ref{eq:fh2fit})) on the
%numerical resolution of the simulations. The left panel shows the
%atomic-to-molecular gas transition as a function of the fitting
%variable $x$ (Equation (\ref{eq:xdef})) for the set of test simulations
%shown in Fig.\ \ref{fig:conv} with the parameter $n_\ast$ fixed to
%$25\dim{cm}^{-3}$. Black squares show the fitting formula (Equation (
%\ref{eq:fh2fit}). In the right panel the parameter $n_*$ is adjusted
%with resolution as $n_*=25(65\dim{pc}/\Delta x)^{1/4}\dim{cm}^{-3}$.}
%\end{figure}

\subsection{Dependence on Averaging Scales}
\label{sec:avgsfl}

\begin{figure}[t]
\includegraphics[scale=0.45]{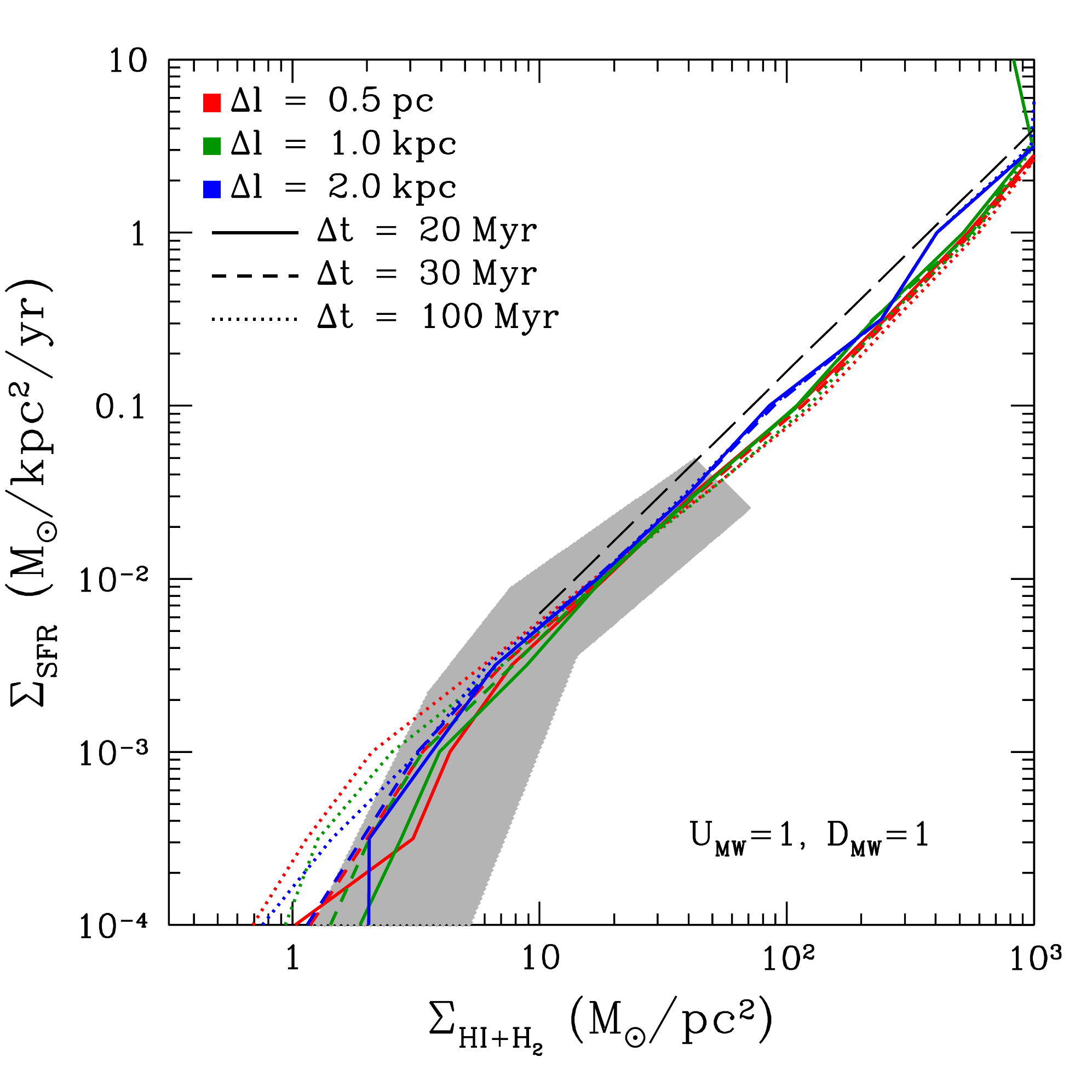}
\includegraphics[scale=0.45]{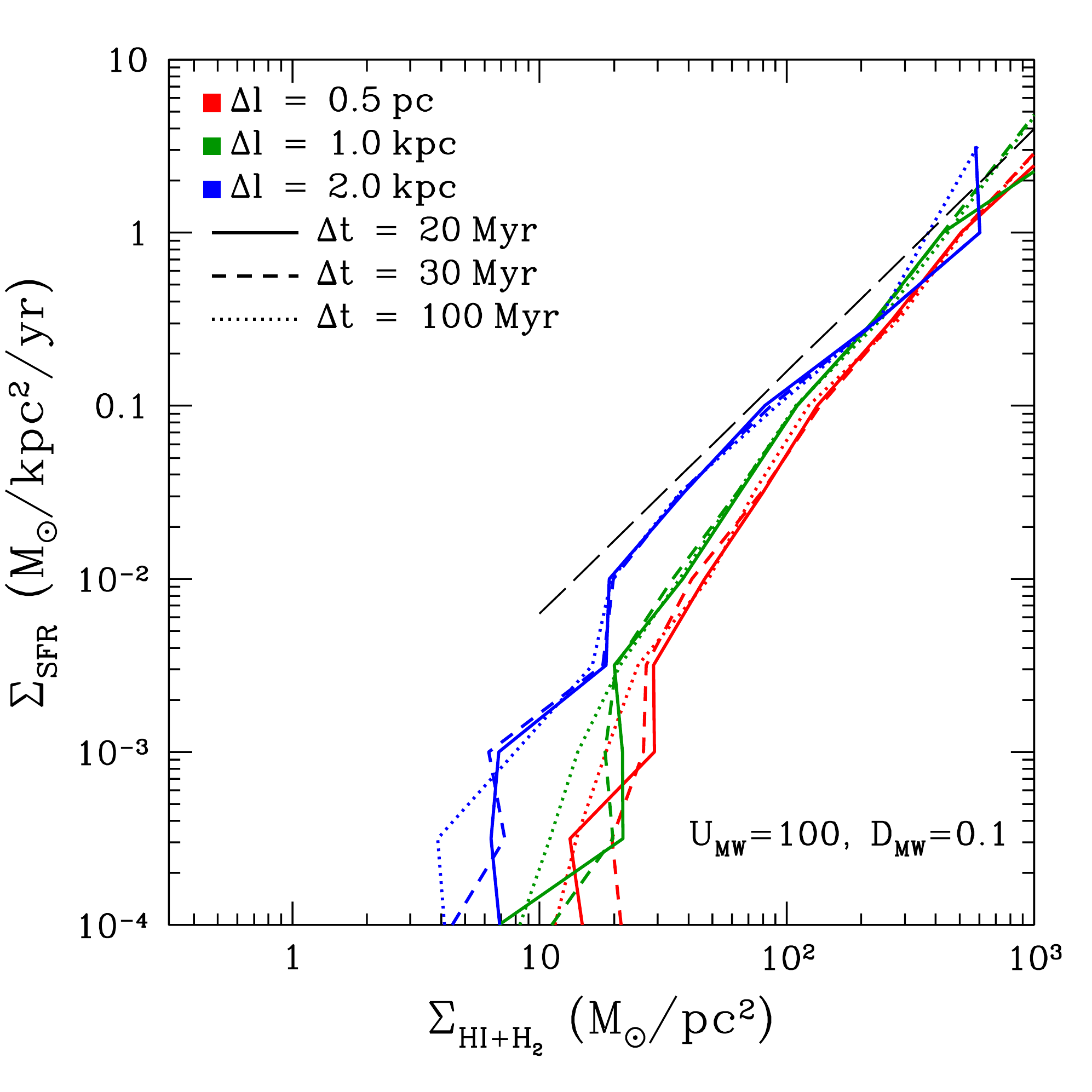}
%\epsscale{1.15}
%\plottwo{\figname{convfh2.ps}}{\figname{convsfl.ps}}
\caption{\label{fig:avgsfl} Dependence of the KS relation on the
spatial and temporal scales over which the star formation rate is
averaged, for two representative sets of parameters: $\U=1,\D=1$ (left
panel) and $\U=100,\D=0.1$ (right panel). Line types and colors show
averaging over spatial scales from $500\dim{pc}$ to $2\dim{kpc}$ and
over time period from $20\dim{Myr}$ to $100\dim{Myr}$.  }
\end{figure}

The exact value of the star formation rate surface density and the gas
surface density in principle can depend on the specific choices for
the spatial and temporal scales over which $\Sgas$ and $\Ssfr$ are
averaged. Observational studies
\citep{sfr:k98a,sfr:srcb07,sfr:blwb08} often use a combination of star
formation estimators that correspond to different temporal
scales. Therefore, the best approach would be to model the
observational methodology exactly, but this is not feasible in
practice. In this paper we adopt a simplified procedure, and select
the fixed values for both the temporal $\Delta t$ and spatial $\Delta
l$ averaging scales. The sensitivity of our results to the exact
choice for these two scales is shown in Figure \ref{fig:avgsfl}. In
general, the KS relations measured in the simulations are robust for
$\Delta t\la 30\dim{Myr}$ and $\Delta l\la 1\dim{kpc}$. For larger
spatial and temporal scales modest trends are
observed. Several processes can contribute to such trends. For
example, if the star formation at low surface densities is
intermittent on the time scale of the averaging (i.e. stars form only
during episods of duration comparable to the averaging time period),
the average $\Ssfr$ can depend on the time period used for
averaging. This may explain the weak trend at low $\Sgas$ with $\Delta
t$. Such trend is also consistent with observations
\citep[e.g.,][]{boissier_etal07}, which show that star formation
derived from the UV flux is more spatially extended compared to the
star formation derived from H$_{\alpha}$, which corresponds to time
period of $\sim 10^7$ years. Overall, our results are quite robust to
changes of spatial and temporal averaging scales within the range of
values used in observations. This relative insensitivity of the KS
relation (besides the weak trends mentioned above) is in general
agreement with observations, which indicate broadly consistent KS
relations derived using different star formation indicators and a wide
range of spatial averaging scales
\citep[e.g.,][]{kennicutt_etal07,sfr:blwb08}.

\bibliographystyle{apj}
\bibliography{ak,ng-bibs/sfr,ng-bibs/self,ng-bibs/sims,ng-bibs/ism,ng-bibs/igm,ng-bibs/hizgal,ng-bibs/misc,ng-bibs/dsh,ng-bibs/atom}

\end{document}